\documentclass[submission, Phys]{SciPost}

\usepackage{graphicx}
\usepackage{dcolumn}
\usepackage{bm}
\usepackage{multirow}

\usepackage{centernot}

\usepackage[normalem]{ulem}

\usepackage{amsmath}
\usepackage{amsthm}
\usepackage{amstext}
\usepackage{amssymb}
\usepackage{mathrsfs}
\usepackage{amsfonts}
\usepackage{amsbsy} 
\usepackage{tensor}
\usepackage{physics}
\usepackage{mathtools}
\usepackage{tikz-cd}
\usepackage{xcolor}

\usepackage[all,matrix,cmtip]{xy}

\usepackage{csquotes}
\MakeOuterQuote{"}

\usepackage{color}
\definecolor{red}{rgb}{1,0,0}
\definecolor{blue}{rgb}{0,0,1}
\definecolor{dblue}{rgb}{0,0,0.4}
\definecolor{green}{rgb}{0,1,0}
\definecolor{black}{rgb}{0,0,0}
\definecolor{white}{rgb}{1,1,1}
\definecolor{niceBlue}{RGB}{20,10,237}
\definecolor{brn}{rgb}{.8,.4,.0}
\definecolor{redo}{rgb}{1,.5,.0}
\definecolor{ddgrn}{rgb}{0,0.4,0}
\definecolor{dgrn}{rgb}{0,0.55,0}
\definecolor{dbl}{rgb}{0,0,0.5}

\usepackage[bbgreekl]{mathbbol}

\newcommand{\Z}{\mathbb{Z}}

\newcommand{\EE}{\mb{E}}
\newcommand{\ssb}{\overset{\mathrm{ssb}}{\longrightarrow}}

\newcommand{\ii}{\hspace{1pt}\mathrm{i}\hspace{1pt}}
\newcommand{\ee}{\hspace{1pt}\mathrm{e}}
\renewcommand{\dd}{\hspace{1pt}\mathrm{d}}

\newcommand{\pp}{\partial}

\newcommand{\bpm}{\begin{pmatrix}}
\newcommand{\epm}{\end{pmatrix}}
\newcommand{\bmm}{\begin{matrix}}
\newcommand{\emm}{\end{matrix}}

\newcommand{\cH}{ {\cal H} }

\usepackage{euscript}

\newcommand{\al}{\alpha} 
 
\newcommand{\del}{\delta} 
\newcommand{\Del}{\Delta}

\newcommand{\ga}{\gamma} 
\newcommand{\Ga}{\Gamma}

\newcommand{\Om}{\Omega} 
\renewcommand{\th}{\theta} 
 
\newcommand{\si}{\sigma} 
\newcommand{\Si}{\Sigma}

\newcommand{\txti}[1]{\textit{#1}}
\renewcommand{\txt}[1]{\text{#1}}

\newcommand{\hstar}{\mathop{*}} 
\def\wdg{{\mathchoice{\,{\scriptstyle\wedge}\,}{{\scriptstyle\wedge}}{{\scriptscriptstyle\wedge}}{{\scriptscriptstyle\wedge}}}} 

\newcommand{\Hom}{\mathrm{Hom}}
\renewcommand{\hom}{\Hom}

\newcommand{\mb}[1]{\mathbb{#1}}

\usepackage{stackrel}

\begin{document}

\begin{center}{\Large \textbf{
Topological aspects of brane fields:\\ solitons and higher-form symmetries
}}\end{center}

\begin{center}
Salvatore D. Pace\textsuperscript{1},
Yu Leon Liu\textsuperscript{2}
\end{center}

\begin{center}
{\bf 1} Department of Physics, Massachusetts Institute of Technology,\\ Cambridge, MA 02139, USA
\\
{\bf 2} Department of Mathematics, Harvard University,\\ Cambridge, MA 02138, USA
\end{center}

\begin{center}
\today
\end{center}

\section*{Abstract}
{\bf
In this note, we classify topological solitons of $n$-brane fields, which are nonlocal fields that describe $n$-dimensional extended objects. We consider a class of $n$-brane fields that formally define a homomorphism from the $n$-fold loop space $\Omega^n X_D$ of spacetime $X_D$ to a space $\mathcal{E}_n$. Examples of such $n$-brane fields are Wilson operators in $n$-form gauge theories. The solitons are singularities of the $n$-brane field, and we classify them using the homotopy theory of ${\EE_n}$-algebras. We find that the classification of codimension ${k+1}$ topological solitons with ${k\geq n}$ can be understood using homotopy groups of $\mathcal{E}_n$. In particular, they are classified by ${\pi_{k-n}(\mathcal{E}_n)}$ when ${n>1}$ and by ${\pi_{k-n}(\mathcal{E}_n)}$ modulo a ${\pi_{1-n}(\mathcal{E}_n)}$ action when ${n=0}$ or ${1}$. However, for ${n>2}$, their classification goes beyond the homotopy groups of $\mathcal{E}_n$ when ${k< n}$, which we explore through examples. We compare this classification to $n$-form $\mathcal{E}_n$ gauge theory. We then apply this classification and consider an ${n}$-form symmetry described by the abelian group ${G^{(n)}}$ that is spontaneously broken to ${H^{(n)}\subset G^{(n)}}$, for which the order parameter characterizing this symmetry breaking pattern is an ${n}$-brane field with target space ${\mathcal{E}_n = G^{(n)}/H^{(n)}}$. We discuss this classification in the context of many examples, both with and without 't Hooft anomalies.
}

\vspace{10pt}
\noindent\rule{\textwidth}{1pt}
\tableofcontents\thispagestyle{fancy}
\noindent\rule{\textwidth}{1pt}
\vspace{10pt}

\section{Introduction}

Excitations in a quantum field theory can be formally described in two drastically different ways. There are the prototypical excitations that are described by the quanta of fields and whose properties are often manifest. However, there are also the excitations described by the topology of the fields, which are called topological solitons\footnote{Topological solitons are instead called topological defects by the solid-state physics community} and whose properties are often much more elusive. Topological solitons appear throughout physics, from condensed matter~\cite{M1979591} to cosmology and particle physics~\cite{VilenkinShellard94}, and play an important role in, for example, continuous phase transitions~\cite{B1971493, B1972610, KT197381, NH197919, SVB0490} and 't Hooft anomalies~\cite{CH8527, HKT191014039, COS191014046, DDK230916749}. While their formal descriptions differ, both excitations describe states in the Hilbert space and are equally important in physical systems. In fact, it is often the case that dualities relate the excitations described by the fields' quanta in one theory to those described by the fields' topology in the dual theory~\cite{MV150505142, KT160601893, SSW160601989}.

The theory and classification of topological solitons of local fields is well established using homotopy groups~\cite{M1979591, VilenkinShellard94}. However, topological solitons of nonlocal fields have yet to receive much attention. In this paper, we classify topological solitons of brane fields. While a local field's quanta correspond to particles, an $n$-brane field is a nonlocal field, and its quanta are $n$-dimensional extended objects. We will always assume that these $n$-dimensional objects do not have a boundary (i.e., they are closed). These extended objects can resemble the ultraviolet (UV) degrees of freedom, like in string theories where the ${n=1}$ brane fields are string fields~\cite{erbin2021string}. However, they may also be emergent and describe the effective infrared (IR) degrees of freedom. In fact, the latter scenario is quite common~\cite{LR152476, Rey339689, PS168191, LW0510, F0607310, BSZ10062267, IM210612610, HK231007993}. For example, while their microscopic degrees of freedom are bosonic particles, superfluid phases in three spatial dimensions host vortex strings in space.

Topological solitons are ubiquitous to spontaneous symmetry breaking phases, where they appear as singularities of the order parameter that characterizes the symmetry breaking pattern. While order parameters of ordinary symmetries are local fields~\cite{LLStatisticalPhysics}, order parameters of $n$-form symmetries are $n$-brane fields~\cite{IM210612610, HK231007993}.\footnote{As $n$-brane fields, they are often parameterized in terms of local $n$-form fields (i.e., Wilson loops, surfaces, etc.). However, this introduces unphysical degrees of freedom through a gauge redundancy, is not always possible (i.e., 't Hooft loops, surfaces, etc.), and is unnatural from the point of view of order parameters in Landau-Ginzburg theory.} An $n$-form symmetry is a generalized symmetry~\cite{M220403045,CD220509545, S230518296, LWW230709215} whose symmetry defects are codimension $(n+1)$ and whose symmetry charges are carried by ${\geq n}$ dimensional objects~\cite{NOc0605316, NOc0702377, GW14125148, W181202517, PW230105261, WZ230312633, BS230402660, BBG230403789, CJ230413751}. Therefore, classifying topological solitons of brane fields also classifies solitons associated with higher-form symmetries. In this paper, we only consider spontaneously broken invertible $n$-form symmetries, which are $n$-form symmetries whose symmetry defects are invertible and described by a group. 

While topological solitons can arise from spontaneously breaking symmetries, their relationship to symmetries is even deeper: they always carry symmetry charges of generalized symmetries~\cite{CT230700939, P230805730, PZB231008554}. These generalized symmetries can be invertible or non-invertible and $0$-form or higher-form symmetries, and their symmetry breaking patterns classify nontrivial disordered phases~\cite{P230805730, PZB231008554}. Therefore, classifying topological solitons of brane fields provides the formal foundation for understanding these generalized symmetries and predicting interesting phase transitions.

The remainder of this paper is as follows. In Sec.~\ref{meanBraneField}, we discuss $n$-brane fields in greater detail. We first further elaborate on \txti{effective} brane fields from the point of view of lattice models. We then discuss the type of brane fields we study in this paper. As we will explain, these are $n$-brane fields that map from the $n$-fold loop space of spacetime while preserving a grouplike $\EE_n$-algebra structure, making them homomorphisms. In Sec.~\ref{sec:ClassDefs}, we derive our classification of topological solitons of $n$-brane fields using the homotopy classes of $\EE_n$-algebra maps (homomorphisms preserving an $\EE_n$-algebra structure). Using techniques from homotopy theory, we work out the classification for a handful of examples. Interestingly, we find that the solitons of $n$-brane fields are equivalent to solitons (i.e., magnetic defects) of $n$-form gauge theory when ${n \leq 2}$, but go beyond those of $n$-form gauge theory when ${n > 2}$. Those exotic solitons are related to the higher homotopy groups of spheres. In Sec.~\ref{applicationToSSBSec}, we apply this classification to order parameters for spontaneously broken higher-form symmetries, discussing examples and the effects of 't Hooft anomalies. Finally, in Sec.~\ref{outlook}, we recap our results and discuss some interesting future directions. Furthermore, the appendices review various topics from homotopy theory used in the main text. In Appendix~\ref{basedVsFreeAppen}, we discuss the relationship between based and free homotopy classes. In Appendix~\ref{LoopSpaceAppendix}, we provide an introduction to iterated loop spaces and $\EE_n$-algebras.

\section{\texorpdfstring{$n$}{n}-brane fields}\label{meanBraneField}

In a quantum many-body theory, a brane field is a quantum field acting on the Hilbert space that creates an extended excitation in space. In ${D=d+1}$ dimensional spacetime, it therefore corresponds to an operator on which these excitations' worldvolumes can end. In this section, we further motivate $n$-brane fields as effective fields starting from lattice models and then discuss their formal description in the continuum.

\subsection{Effective brane fields from lattice models}\label{effBraneLattice}

To further motivate effective $n$-brane fields, let us consider a $d$-dimensional lattice theory. We assume that the Hilbert space $\cH$ has a tensor product decomposition ${\cH = \bigotimes_{i}\cH_{i}}$, where $i$ can label vertices, links, plaquettes, etc of the lattice. We assume that there exists a local Hamiltonian ${H = \sum_{j} H_{j}}$ that respects the lattice symmetries and governs the system. Here, $j$ labels vertices, links, plaquettes, etc of the lattice, but they need not be the same as those labeled by $i$. The precise details of $\cH_i$ and $H_{j}$ are unimportant for the following discussion.

Given the microscopic theory $H$, it is commonplace to investigate its behavior at long distances/low energy. This is typically done by constructing an effective IR continuum field theory. A standard technique to connect a quantum lattice Hamiltonian to a continuum field theory is to perform a mean-field theory analysis. Starting with a state ${\ket{\{\psi_i\}}\in \cH}$ (e.g., a coherent state), where each mean-field parameter ${\psi_{i}\in\mathbb{C}}$ is assumed to be time-dependent, the lattice phase-space Lagrangian is defined as ${L(\psi_{i}) \equiv \bra{\{\psi_i\}} \ii\frac{\dd}{\dd t}-H\ket{\{\psi_i\}}}$. Upon coarse-graining the lattice, $L(\psi_{i})$ becomes the continuum phase-space Lagrangian meant to describe the IR.  

How successful this continuum theory is depends on how accurately the state ${\ket{\{\psi_i\}}}$ reflects the Hamiltonian's IR physics. For example, it is typically assumed that $i$ are only vertices of the lattice and that upon coarse-graining, ${\{\psi_{i}(t)\}}$ becomes the local field $\psi(x)$. However, regardless of how ${\ket{\{\psi_i\}}}$ is then chosen, there is already an implicit assumption regarding the IR of $H$ in doing this. Indeed, we assumed that ${\{\psi_{i}\}}$ coarse grains to the local field $\psi(x)$. This is not always the case because the IR degrees of freedom are not required to be particles. For example, if $H$ has topological order, the IR degrees of freedom are extended objects~\cite{LW0510} and the IR effective field theory will be a topological field theory (which has no local degrees of freedom whatsoever).

Suppose we knew the IR degrees of freedom were loops (i.e., 1-branes). It is natural to then interpret the labels $i$ as the links $\ell$ of the lattice so the UV degrees of freedom are viewed as small open strings. The relationship between the UV and IR degrees of freedom can then be understood as the IR loops being formed by the small UV open strings. This scenario precisely happens in certain quantum spin liquids~\cite{SB160103742}, where the UV degrees of freedom are spins on a lattice that form loops at low energies by pointing head to tail with nearest neighbor spins. Therefore, taking into account the IR degrees of freedom, coarse-graining ${\{\psi_{i}\}}$ gives rise to a string field $\psi(\ga)$ and not a local field.

Generalizing this from loops to $n$-branes is straightforward. With the foresight of the IR degrees of freedom being $n$-branes, we view the UV degrees of freedom  as residing on $n$-cells of the lattice. Here, an $n$-cell refers to an $n$-dimensional component of the lattice (i.e., $0$-cells are lattice sites, $1$-cells are lattice links, $2$-cells are plaquettes, etc.). In the IR, these $n$-cells form $n$-branes and the mean-field parameters $\{\psi_{i}\}$ coarse grain to an $n$-brane field.

To summarize, $n$-branes can emerge in the IR even when they are absent in the UV. In this paper, when we discuss $n$-brane fields, we will make no assumptions whether they are elementary or emergent degrees of freedom.

\subsection{Homomorphism \texorpdfstring{$n$}{n}-brane fields}\label{BraneFieldDef}

In this paper, we study the topological, kinematic aspects of $n$-brane fields rather than their dynamics. Probing the dynamics would require constructing actions in terms of $n$-brane fields. This is interesting but not needed for classifying topological solitons. We refer the reader to Refs.~\citenum{IM210612610, HK231007993}, which are two recent papers that constructed actions for $n=1$ and general $n$, respectively, in the context of higher-form symmetries (see Sec.~\ref{applicationToSSBSec}).

One perspective~\cite{HK231007993} of an $n$-brane field is a field $\Psi$ that assigns a value in a space $T$ to an $n$-brane. In this view, an $n$-brane is formally a pair $(M_n, f)$ consisting of an $n$-dimensional manifold $M_n$ together with all maps ${f \colon M_n \to X_D}$. Denoting the space of all such $n$-branes in $X_D$ by $\txt{Brane}_n(X_D)$, this $n$-brane field is a map
\begin{equation}\label{bBraneFieldDef}
    \Psi\colon \txt{Brane}_n(X_D)\to T.
\end{equation}
While $\Psi$ is arguably the most general construction of an $n$-brane field, it lacks the structure to describe two $n$-branes being fused together. This can be enforced dynamically by including a term in the $n$-brane field theory action (e.g., Eq.~27 in Ref.~\citenum{IM210612610} for ${n=1}$). However, deriving physical consequences using this action is challenging. 

In this paper, we work with a different definition of $n$-brane fields that encode this fusion structure kinematically instead. As we will now describe, our definition of $n$-brane field $\psi$ makes it a homomorphism respecting this fusion structure. Furthermore, we consider only $n$-fold loops as $n$-branes, and not the general $n$-brane ${(M_n,f)}$ described above.

Before stating the general definition of $\psi$, let us first discuss the case where ${n=1}$ (i.e., string fields). We assume that spacetime $X_D$ has a based point $x\in X_D$ and consider a $1$-brane to be the based loop
\begin{equation}\label{based1FoldLoop}
    \ga_1\colon (S^1,s)\to (X_D,x),
\end{equation}
where $s$ is the based point of $S^1$. The space of loops~\eqref{based1FoldLoop} is called the loop space $\Omega X_D$ of $X_D$. Unlike ${\txt{Brane}_1(X_D)}$ defined above, $\Omega X_D$ includes the multiplication structure encoding the concatenation of two loops. Indeed, given two based loops ${\ga^{(1)},\ga^{(2)}\in \Om X_D}$, we can concatenate them to give a new based loop ${\ga^{(1)}\circ\ga^{(2)}\in \Om X_D}$. This defines a homotopically associative (but not commutative) multiplication on $\Om X_D$:
\begin{equation}
    \circ\colon \Om X_D\times\Om X_D\to \Om X_D.
\end{equation}  
The loop concatenation structure upgrades $\Omega X_D$ from a based space to what is called an $\EE_1$-algebra. Furthermore, $\Om X_D$ is a \emph{grouplike} $\EE_1$-algebra because every loop has an inverse with respect to the concatenation product given by the loop running in the opposite direction. 

Since our $1$-brane field $\psi$ is defined as a homomorphism respecting  loop concatenation, there is a multiplication operation $\bm{\cdot}$ for which
\begin{equation}\label{phib1brane}
    \psi(\ga_n^{(1)})\bm{\cdot} \psi(\ga_n^{(2)}) = \psi(\ga_n^{(1)}\circ \ga_n^{(2)}).
\end{equation}
This implies that the target space $\mathcal{E}_1$ of $\psi$ has a grouplike $\EE_1$-algebra structure whose multiplication operation is $\bm{\cdot}$, and  $\psi$ is a map of $\EE_1$-algebras
\begin{equation}\label{based1Branefield}    
    \psi\colon\Om^1 X_D\to \mathcal{E}_1.
\end{equation}

We now generalize our definition to arbitrary $n$. We define an $n$-brane field $\psi$ as a field that inputs an $n$-fold loop and respects the $n$-fold loop concatenation structure.\footnote{We refer the reader to appendix~\ref{LoopSpaceAppendix} for a detailed introduction to $n$-fold loop spaces (\ref{itLoopSpaceIntro}), $n$-fold loop concatenation and $\EE_n$-algebras (\ref{appendixEn}), and maps of $\EE_n$-algebras (\ref{mapsOfEnAlgs}).} An $n$-fold loop $\ga_n$ is defined as a based map 
\begin{equation}\label{nFoldLoopDef}
    \ga_n\colon (S^n,s)\to (X_D,x),
\end{equation}
where $s$ denotes the basepoint of the $n$-sphere $S^n$. The space of all $n$-fold loops in $(X_D,x)$ is called the $n$-fold loop space $\Om^n X_D$ of $X_D$. For ${n>0}$, $\Om^n X_D$ has the mathematical structure of a grouplike $\EE_n$-algebra. $\EE_n$-algebras describe the fusion of points in $I^n = [0,1]^n$ and are called grouplike when each point has an inverse under fusion.\footnote{Grouplike $\EE_n$-algebras are homotopic generalizations of groups, where the commutativity and associativity only hold up to homotopy. For example, groups are grouplike $\EE_1$-algebras and abelian groups are grouplike $\EE_\infty$-algebras.} Since $\psi$ is defined as a homomorphism respecting $n$-fold loop concatenation, there is a multiplication operation $\bm{\cdot}$ for which
\begin{equation}\label{phibnbrane}
    \psi(\ga_n^{(1)})\bm{\cdot} \psi(\ga_n^{(2)}) = \psi(\ga_n^{(1)}\circ \ga_n^{(2)}).
\end{equation}
This implies that the target space $\mathcal{E}_n$ of $\psi$ has a grouplike $\EE_n$-algebra structure whose multiplication operation is $\bm{\cdot}$, and  $\psi$ is a map of $\EE_n$-algebras
\begin{equation}\label{basednBranefield}    
    \psi\colon\Om^n X_D\to \mathcal{E}_n.
\end{equation}
To distinguish $\psi$ with Eq.~\eqref{bBraneFieldDef}, one may call $\psi$ a homomorphism $n$-brane field. However, when it is clear from the context, we will often just call them $n$-brane fields.


When ${n=1}$, the primary difference between definitions~\eqref{bBraneFieldDef} and~\eqref{basednBranefield} is the kinematic inclusion of loop concatenation. For ${n>1}$, however, there is an important physical difference as well. In particular, $\txt{Brane}_n(X_D)$ in~\eqref{bBraneFieldDef} includes $n$-branes of various topologies while $\Om^n X_D$ in~\eqref{basednBranefield} only includes those homotopic to an $n$-sphere. For example, when ${n=2}$, the homomorphism $n$-branes in ~\eqref{basednBranefield} do not include $2$-branes that are homotopic to a torus, only those homotopic to a 2-sphere. 

We remark that definition~\eqref{basednBranefield} of an $n$-brane field makes sense when ${n=0}$. While an $\EE_0$-algebra has no multiplication structure, it still coincides with the notion of based spaces. Indeed, a ${0}$-fold loop is determined only by where the non-basepoint of $S^0$ goes, and so ${\Om^0 X_D = (X_D,x)}$. Therefore, a $0$-brane field is just a local field mapping from the based space $(X_D,x)$ to a target space $\mathcal{E}_0$ with a based point. This makes the definition~\eqref{basednBranefield} of $n$-brane fields a natural generalization of based local fields. As we will see in Sec.~\ref{localFieldsDefects}, based local fields are commonly considered when classifying topological solitons in nonlinear $\si$-models. Similarly, homomorphism $n$-branes will classify topological solitons in $n$-brane theories.

\subsection{\texorpdfstring{$n$}{n}-brane fields versus \texorpdfstring{$n$}{n}-form gauge fields}\label{braneVsGauge}

A common class of theories whose physical degrees of freedom are described by $n$-brane fields are pure $n$-form gauge theories. An $n$-form $G$ gauge theory is a theory whose gauge fields are locally differential $n$-forms of spacetime. When ${n=1}$ this is ordinary gauge theory, whose gauge field $a^{(1)}$ is mathematically a connection on a principal $G$-bundle whose isomorphism classes are in one-to-one correspondence with the homotopy classes of maps from $X_D$ to ${BG}$. More generally, the physical information of a $n$-form $G$ gauge field $a^{(n)}$, where $G$ is abelian for ${n>1}$, is given by the homotopy classes of maps from $X_D$ to ${B^n G}$. \footnote{together with differential cohomology data when $G$ is not discrete.} 

Given an $n$-form gauge field $a^{(n)}$ with abelian group $G$, we have the Wilson operator\footnote{From the point of view of higher-form symmetries, $W_a(C_n)$ can be thought of as the Goldstone parameterization of the order parameter $n$-brane field~\cite{IM210612610, HK231007993}.}
\begin{equation}\label{WilsonBrane}
    W_a(C_n) = \exp[\ii\int_{C_n}a^{(n)}].
\end{equation}
This defines an $n$-brane field
\begin{equation}
    W_a(-) \colon \txt{Brane}_n(X_D) \to G
\end{equation}
associated to $a^{(n)}$. For example, if ${G_n = U(1)}$, $a^{(n)}$ is a $U(1)$ $n$-form gauge field and the target space of $W_a(C_n)$ is $U(1)$. Another example is ${G = \Z_N}$, in which case $a^{(n)}$ can be a ${U(1)}$ $n$-form gauge field satisfying ${\dd a = 0}$ and the quantization condition ${\frac{N}{2\pi}\int_{C_n} a^{(n)} \in \Z}$. Then, in this example, $W_a(C_n)$ is an $n$-brane field with target space $\Z_N$. 

We note that $C_n$ does not have to be embedded in $X_D$ (i.e., a $n$-submanifold of $X_D$). Even from the lattice point of view advocated in Sec.~\ref{effBraneLattice}, restricting to embeddings is unnecessary. Indeed, a general $n$-fold loop $\ga_n$ can be realized on a lattice, including those that are not injective which can be done by constructing the lattice and $n$-fold loops using simplicial sets (e.g., Ref.~\cite{MS1052000}). See, for instance, Ref.~\cite{PZB231008554} for a discussion of lattice models and simplicial sets.

The Wilson operator satisfies
\begin{equation}
    W_a(C^{(1)}_n)W_a(C^{(2)}_n) = W_a(C^{(1)}_n \cup C^{(2)}_n).
\end{equation}
This hints at a relation between $n$-form gauge fields and homomorphism $n$-brane. Indeed, restricting $C_n$ to $n$-fold loops, the Wilson operator Eq.~\eqref{WilsonBrane} defines an $\EE_n$-homomorphism ${\Omega^n X_D \to G}$ and thus a homomorphism $n$-brane field with $\mathcal{E}_n = G$.\footnote{$G$ is an abelian group, thus a grouplike $\EE_\infty$-algebra. Any $\EE_n$-algebra is an $\EE_m$-algebra for $0 \leq m \leq n \leq \infty$.} Mathematically, this relation arises because any map of based spaces ${f \colon X \to Y}$ introduces a $\EE_n$-algebra map ${\Omega^n f \colon \Omega^n X \to \Omega^n Y}$ (see Appendix.~\ref{deloopAppendix}). In our case, ${Y = B^n G}$ and ${\Omega^n B^n G \simeq G}$.
Therefore, the Wilson operator takes a $n$-form $G$ gauge field to a homomorphism $n$-brane with target space $\mathcal{E}_n = G$. It can, therefore, be viewed as a map
\begin{equation}
    W_{-}(C_n)\colon n\txt{-form gauge fields}\to \txt{homomorphism }n\txt{-brane fields}.
\end{equation}

Let us make a few remarks. Firstly, the map ${W_{-}(C_n)}$ is not injective or surjective. Although $n$-form gauge fields are determined by their holonomies over \textit{all} possible $n$-submanifold of spacetime, since homomorphism $n$-branes only include $n$-spheres, ${W_{-}(C_n)}$ can map two different $n$-form gauge fields to the same homomorphism $n$-brane fields. Furthermore, not all homomorphism $n$-brane fields can come from a $n$-form gauge field. For example, while there are no non-trivial $3$-form gauge fields on $S^2$ as ${H^3(S^2; \Z) = 0}$, there are non-trivial homomorphism $3$-brane fields as ${\Omega^3 S^2  \simeq \mathbb{Z}}$. The discrepancy is generally related to the higher homotopy groups of spheres. Consequently, as we will see in Sec.~\ref{braneSolvsGaugeSol}, the topological solitons for homomorphism $n$-brane will include those present in $n$-form gauge theory, but can include additional topological solitons that are beyond gauge theory.

\section{Classification of topological solitons}\label{sec:ClassDefs}

In this section, we present our classification of topological solitons of the (homomorphism) $n$-brane fields defined in Sec.~\ref{BraneFieldDef}. When discussing the dimensionality of a topological soliton, we will always refer to its codimension in spacetime. So, a codimension $k$ topological soliton in $D$ dimensional spacetime is ${(D-k)}$-dimensional.  We remind the reader of the typical terminology, where codimension $1$ solitons are called domain walls, codimension $2$ solitons are vortices, codimension $3$ solitons are hedgehogs/monopoles, and codimension $D$ solitons are instantons.

\subsection{Reviewing solitons of local fields}\label{localFieldsDefects}

Before discussing topological solitons of brane fields, we first review their classification for a local field
\begin{equation}
    \phi\colon X_D\to T,
\end{equation}
where we assume that $T$ is path-connected. Suppose there is a codimension ${(k+1)}$ topological soliton whose core is a codimension ${(k+1)}$ submanifold $\Delta$. The local field $\phi$ is only well-defined away from the core. Therefore, in the presence of the topological soliton, the local field is a map
\begin{equation}
    \phi^\Del\colon X_D-\Del\to T.
\end{equation}
The topological soliton manifests as a singularity in the field $\phi^\Del$.

It is useful to view topological solitons from the ``detection'' point of view. That is, we use a closed $k$-submanifold $\Si_k$ of ${X_D-\Del}$ to detect the topological soliton. The local field $\phi^\Del$ restricts to a map 
\begin{equation}\label{localFieldRestrict}
    \phi^\Del|_{\Si_k}\colon\Si_k\to T.
\end{equation}
Since topological solitons are characterized by the topology of $T$, two maps ${\phi^\Del|_{\Si_k}}$ that can be continuously deformed from one to another detect equivalent topological solitons using $\Si_k$. In particular, if ${\phi^\Del|_{\Si_k}}$ is homotopic to a constant map, then $\Si_k$ detects no solitons. On the other hand, when the free homotopy class $[\phi^\Del|_{\Si_k}]_f$ is a nontrivial element of ${[\Si_k, T]_f}$,\footnote{The notation ${[A,B]_f}$ denotes the set of free maps from ${A}$ to ${B}$ up to free homotopy.} then $\Si_k$ detects a non-trivial topological soliton~\cite{M1979591}. We see that topological solitons detected by $\Si_k$ are classified by ${[\Si_k, T]_f}$.

To locally identify topological solitons, we pick a point ${d \in \Delta}$ and consider the linking sphere ${\Si_k = S^{k}}$ to it. Then, the free homotopy class $[\phi|_{\Si_k}] \in { [S^k, T]_f}$ being nontrivial implies that we cannot extend $\phi$ across $d$, and there exists a topological soliton at the connected component of ${d \in \Delta}$. Since this topological soliton is detected by linking with $S^k$, it is codimension ${k+1}$. Therefore, codimension ${k+1}$ solitons are classified by
\begin{equation}\label{freeHomotopyClassesT}
    [S^k, T]_\txt{f}.
\end{equation}
We emphasize that this is the set of free homotopy classes of maps from $S^k$ to $T$. They are generally not the same as the homotopy groups of $T$, which are the based homotopy classes of based maps (see Appendix~\ref{basedVsFreeAppen}). However, since based maps and based homotopy classes are often easier to work with, it is worthwhile to reformulate this classification in terms of based homotopy.

To understand the relation between free and based homotopy classes, we must understand how changing the basepoint affects the based homotopy class. Let us choose basepoints ${s\in S^k}$ and ${t\in T}$ and consider the based maps from ${(S^k,s)}$ to ${(T,t)}$ modulo based homotopy, denoted by ${[(S^k,s),(T,t)]_\txt{b}}$. Given a path from ${t\in T}$ to ${t'\in T}$, viewed as a continuous change of basepoints, we can transport a based map to $(T, t)$ to a based map to $(T, t')$. This induces an action of paths on based homotopy classes. In particular, it restricts to an action on $\pi_1(T,b)$ on $ [(S^k, s), (T,t)]_{\txt{b}}$. In the case that $T$ is path-connected, it turns out that the free homotopy classes are the same as the based homotopy classes modulo this $\pi_1(T,t)$ action:
\begin{equation}\label{eq:based-homotopy-to-free}
    [S^k,T]_{\txt{f}} = [(S^k, s), (T,t)]_{\txt{b}}/\pi_1(T, t).
\end{equation}
See Appendix~\ref{basedVsFreeAppen} for a  derivation of this equation. Since the set of based homotopy classes $[(S^k, s), (T,t)]_{\txt{b}}$ is equal to the $k$-th homotopy group $\pi_k(T,t)$, Eq.~\eqref{eq:based-homotopy-to-free} becomes \footnote{When it causes no confusion, we will sometimes suppress the basepoint and denote $\pi_k(T, t)$ as just $\pi_k(T)$. }
\begin{equation}\label{eq:based-homotopy-classify-defects}
    [S^k,T]_{\txt{f}} = \pi_k(T)/\pi_1(T).
\end{equation}

From the above, we see that codimension $(k+1)$ topological solitons are classified by the free homotopy classes~\eqref{freeHomotopyClassesT}, which is equivalently $\pi_k(T)$ modulo the $\pi_1(T)$ action. A benefit of understanding the classification using homotopy groups is that it provides information on the fusion of topological solitons. Indeed, recall from Sec.~\ref{BraneFieldDef} that the space of based maps ${S^k \to T}$ has a multiplication structure by $k$-fold loop concatenation. In this context, this multiplication physically corresponds to fusing topological solitons. The classification~\eqref{eq:based-homotopy-classify-defects} means that fusing topological solitons labeled by elements of $\pi_k(T)$ can have multiple outcomes that fall into the orbit of the $\pi_1(T)$ action on $\pi_k(T)$. We emphasize, however, that this fusion structure arose only when we fixed a basepoint. Correctly capturing this formal structure will be crucial when generalizing to $n$-branes in the next subsection.

Finally, in anticipation of Sec.~\ref{nBraneDefects}, let us translate the classification~\eqref{eq:based-homotopy-classify-defects} into the language of $\EE_n$-algebras. We introduced $\EE_n$-algebras in Sec.~\ref{BraneFieldDef}, and here we only need to use the fact that an $\EE_0$-algebra is a space with a basepoint. Therefore, the based spaces ${(S^k,s)}$ and ${(T,t)}$ can be viewed as the $\EE_0$-algebras ${\Om^0 S^k}$ and $\mathcal{E}_0$, respectively, and the based map from ${\Om^0S^k = (S^k,s)}$ to ${\mathcal{E}_0 = (T,t)}$ is a map of $\EE_0$-algebras. Furthermore, in terms of $\EE_0$-algebras, the based homotopy classes ${[(S^k,s), (T,t)]_{\txt{b}}}$ become the homotopy classes of $\EE_0$ maps
\begin{equation}\label{0FormBased}
    [\Om^0S^k,\mathcal{E}_0]_{\EE_0} \equiv [(S^k,s), (T,t)]_{\txt{b}} = \pi_k(T).
\end{equation}
Using this language, the classification~\eqref{eq:based-homotopy-classify-defects} becomes
\begin{equation}\label{E0algClass}
    [S^k,T]_{\txt{f}} = [\Om^0 S^k,\mathcal{E}_0]_{\EE_0}/[\Om^0 S^1,\mathcal{E}_0]_{\EE_0}.
\end{equation}

\subsection{Solitons of \texorpdfstring{$n$}{n}-brane fields}\label{nBraneDefects}

Having reviewed the classification of topological solitons of local fields, let us now classify topological solitons of $n$-brane fields $\psi$ defined by Eq.~\eqref{basednBranefield}. Given a topological soliton whose core is the submanifold $\Del$ and the $n$-brane field ${\psi^\Del\colon \Om^n(X_D-\Del)\to \mathcal{E}_n}$, we detect the soliton using the $k$-submanifold $\Si_k$ of ${X_D-\Del}$ and restricting $\psi^\Del$ to the $\EE_n$-algebra map 
\begin{equation}
\psi^\Del|_{\Si_k}\colon \Omega^n \Si_k \to \mathcal{E}_n.
\end{equation}
We remark that $\Omega^n \Si_k$ is a subalgebra of ${\Om^n(X_D-\Del)}$. As for local fields, topological solitons detected by $\Si_k$ are characterized by the homotopy classes of $\psi^\Del|_{\Si_k}$. However, now $\psi^\Del|_{\Si_k}$ is an $\EE_n$-algebra map, and therefore we consider $\EE_n$-algebra homotopies between $\EE_n$-algebra maps. 
We denote the homotopy classes of $\EE_n$-algebras as ${[\Omega^n \Si_k, \mathcal{E}_n]_{\EE_n}}$. For example, these are based homotopy classes of based maps when ${n = 0}$ and homotopy classes of group homomorphisms when ${n=1}$. In the case that the $k$-submanifold ${\Si_k = S^k}$ is the linking sphere to $\Delta$, the homotopy class \begin{equation}\label{loopSphereOP}
    [\Om^n S^k,\mathcal{E}_n]_{\EE_n}
\end{equation}
locally detects codimension ${(k+1)}$-topological solitons. Notice that setting ${n=0}$ reproduces Eq.~\eqref{0FormBased}, and like for local fields, Eq.~\eqref{loopSphereOP} describes the local fusion of codimension ${(k+1)}$ topological solitons.

Eq.~\eqref{loopSphereOP} can be simplified into a more convenient expression using the duality between iterated loop spaces and delooping.\footnote{We review delooping and its relation to iterated loop spaces in detail in Appendices~\ref{deloopAppendix} and~\ref{loopDeloopDuality}, respectively.} Whereas looping $\Om$ takes an $\EE_m$-algebra to an $\EE_{m+1}$-algebra, the delooping operation $B$ is an inverse construction that takes an $\EE_m$-algebra to an $\EE_{m-1}$-algebra. Using the delooping operation, the homotopy classes~\eqref{loopSphereOP} can be rewritten as
\begin{equation}\label{loopSphereOP2}
    [\Om^n S^k,\mathcal{E}_n]_{\EE_n} = [B^n\Om^n S^k,B^n\mathcal{E}_n]_{\EE_0},
\end{equation}
where $B^n$ is shorthand for acting $B$ $n$-times. Notice that this expresses the homotopy classes of $\EE_n$-algebra maps as based homotopy classes of based maps from ${B^n\Om^n S^k}$ to ${B^n\mathcal{E}_n}$, which are both $\EE_0$-algebras (based spaces). 

For any $\EE_0$ algebra $X$, ${B^n \Omega^n X \simeq X}$ if ${\pi_i(X) = 0}$ for all ${i < n}$. Therefore, for ${k\geq n}$, ${B^n \Om^n S^k \simeq S^k}$ because ${\pi_i(S^k) = 0}$ for ${i<k}$, and Eq.~\eqref{loopSphereOP2} simplifies to
\begin{equation}\label{loopSphereOP3}
[\Om^n S^k,\mathcal{E}_n]_{\EE_n} = [S^k ,B^n\mathcal{E}_n]_{\EE_0} = \pi_k (B^n\mathcal{E}_n) = \pi_{k-n}(\mathcal{E}_n)\quad\quad (k\geq n),
\end{equation}
where the last step follows from the properties of $B$. So, for ${k\geq n}$, we find that Eq.~\eqref{loopSphereOP} can be understood in terms of homotopy groups of $\mathcal{E}_n$. However, for ${k < n}$, the space $B^n \Om^n S^k$ is generally not homotopic equivalent to $S^k$. $B^n \Om^n S^k$ is the $(n-1)$-th stage of the Whitehead tower of $S^k$ and satisfies
\begin{equation}
    \pi_i(B^n\Om^nS^k) = \begin{cases}
        0\quad\quad &i < n,\\
        \pi_i(S^k)\quad\quad &i \geq n.
    \end{cases}
\end{equation}
When $k>1$, $B^n\Om^nS^k$ is a nontrivial space for arbitrary $n$.

Like local fields, Eq.~\eqref{loopSphereOP3} fails to provide a classification since it depends on the unphysical choice of basepoint ($n$-fold loops are based loops). For local fields, moving the basepoint by deforming $\Si_k$ defined an action of ${[\phi^\Del_\txt{b}|_{S^1}]\in [(S^1,s),(T,t)]_\txt{b} = \pi_1(T)}$ on $\pi_k(T)$. For $n$-brane fields, we can again deform the basepoint to form a loop ${\gamma \colon S^1 \to X_D}$, but now $\psi^\Del$ restricts to an element ${\psi^\Del|_{S^1} \colon \Omega^n S^1\to \mathcal{E}_n}$ and defines an action of ${[\Omega^n S^1, \mathcal{E}_n]_{\EE_n}}$. Therefore, the topological solitons of an $n$-brane field detected by $\Si_k$ are classified by
\begin{equation}\label{nBraneSolClass}
        [\Omega^n \Si_k, \mathcal{E}_n]_{\EE_n}/[\Omega^n S^1, \mathcal{E}_n]_{\EE_n} = [\Omega^n \Si_k, \mathcal{E}_n]_{\EE_n}/\pi_1(B^n \mathcal{E}_n),
\end{equation}
where we used ${[\Omega^n S^1, \mathcal{E}_n]_{\EE_n} = \pi_1(B^n \mathcal{E}_n)}$. Note that setting ${n=0}$ recovers the classification for local fields~\eqref{E0algClass}. Since ${[\Omega^n S^1, \mathcal{E}_n]_{\EE_n} = 0}$ for ${n > 1}$, the ${[\Omega^n S^1, \mathcal{E}_n]_{\EE_n}}$ action matters only for ${n= 0,1}$. Eq.~\eqref{nBraneSolClass} classifies codimension ${k+1}$ solitons when $\Si_k$ is the linking sphere $S^k$. Furthermore, when ${k\geq n}$, it simplifies to
\begin{equation}\label{nBraneSolClass2}
        [\Omega^n S^k, \mathcal{E}_n]_{\EE_n}/[\Omega^n S^1, \mathcal{E}_n]_{\EE_n} = \pi_{k-n}(\mathcal{E}_n)/\pi_{1-n}(\mathcal{E}_n)\quad\quad (k\geq n),
\end{equation}
where ${\pi_{1-n}(\mathcal{E}_n)}$ is trivial when ${1-n < 0}$. Therefore, for $n$-brane fields, the $\pi_1(B^n \mathcal{E}_n)$ action is only nontrivial when ${n=1}$, in which case ${[\Omega^1 S^k, \mathcal{E}_1]_{\EE_1} = \pi_{k}(B\mathcal{E}_1)}$ for all $k$.

From Eq.~\eqref{nBraneSolClass2}, we find the classification of topological solitons when ${k\geq n}$ is in terms of only the homotopy groups of $\mathcal{E}_n$. Interestingly, when ${k<n}$, the classification goes beyond ${\pi_k(\mathcal{E}_n)}$ and is generally nonzero. In this case, the $n$-brane detects a codimension ${k+1}$ topological soliton by collapsing the $n$-sphere onto a ${k<n}$ dimensional submanifold of spacetime. For example, a $3$-brane that detects a codimension $3$ soliton collapses the $3$-sphere onto a $2$-sphere via the Hopf map.

\subsection{Comparison to gauge theory}\label{braneSolvsGaugeSol}

As mentioned in Sec.~\ref{braneVsGauge}, some $n$-brane fields $\psi$ correspond to the Wilson operators in abelian $\mathcal{E}_n$ $n$-form gauge theory. Therefore, it is natural to expect that some topological solitons of $\psi$ correspond to the ``topological solitons'' (the magnetic defects) of $\mathcal{E}_n$ $n$-form gauge theory. Codimension ${k+1}$ topological solitons of $\mathcal{E}_n$ $n$-form gauge theory are classified by
\begin{equation}\label{nformSol}
    H^{n}(S^k,\mathcal{E}_n) \equiv [S^k,B^n\mathcal{E}_n] = \pi_k(B^n\mathcal{E}_n)/\pi_1(B^n\mathcal{E}_n).
\end{equation}
For example, ${H^{1}(S^2,U(1)) = H^{2}(S^2,\Z) = \Z}$ classifies the 't Hooft lines in $U(1)$ gauge theory while ${H^{1}(S^1,\Z_N) = \Z_N}$ classifies the magnetic defects of $\Z_N$ gauge theory. Note that there are no gauge theory solitons when $k < n$ because $B^n\mathcal{E}_n$ is $n$-connected.

In the previous subsection, we found that the classification of an $n$-brane field's topological solitons matched Eq.~\eqref{nformSol} when ${k\geq n}$ (see Eq.~\eqref{loopSphereOP3}). For ${k < n}$, their classification was given by the general result ${[\Om^n S^k,\mathcal{E}_n]_{\EE_n}}$, which can be non-zero (see Sec.~\ref{exBraneSol} for examples). Hence, codimension ${k+1}$ topological solitons of $n$-brane fields with ${k<n}$ go beyond gauge theory.

Here, we will make two brief remarks comparing this further to the gauge theory classification.

Firstly, we note that the $n$-brane classification matches the $n$-form gauge theory classification when ${n=1}$ and ${n=2}$. This is because ${[\Om^n S^k,\mathcal{E}_n]_{\EE_n}}$ is trivial when ${k < n}$ in both of these cases, so all that is left is $k\geq n$ which matches the gauge theory classification. The reason why ${[\Om^n S^k,\mathcal{E}_n]_{\EE_n}}$ is trivial when ${k < n}$ is because ${\Om^n S^k}$ is homotopically trivial since there are no nontrivial ways to map $n$ spheres into ${k<n}$ sphere for ${n=1,2}$. The first $n$ for which the $n$-brane fields' topological solitons go beyond $n$-form gauge theory is ${n=3}$.

Secondly, when ${[\Om^n S^k,\mathcal{E}_n]_{\EE_n}}$ is nontrivial, it involves a nontrivial mapping of a $n$-sphere onto a $k$-sphere with ${k<n}$. Here, nontrivial means it corresponds to a nontrivial element of $\pi_n(S^k)$. However, since all $n$-form gauge field on $S^k$ are trivial in this case(i.e., ${H^n(S^k, \mathcal{E}_n) = 0}$ when ${k<n}$), any such nontrivial ${S^n\to S^k}$ map pulls back to a trivial $n$-form gauge field on $S^n$. Therefore, these $n$-branes do not correspond to, and go beyond, $n$-form gauge fields.

\subsection{Examples}\label{exBraneSol}

Having worked out the general classification of topological solitons for $n$-brane fields, let's now consider some examples for different $\mathcal{E}_n$. We will consider similar examples in Sec.~\ref{applicationToSSBSec} in the context of spontaneously broken higher-form symmetries.

We first consider the case where $\mathcal{E}_n$ is discrete, which we emphasize by denoting it as $\mathcal{E}^{\txt{discrete}}_{n}$. In this case, Eq.~\eqref{loopSphereOP} can be simplified for general $n$ and $k$. Since $\mathcal{E}^{\txt{discrete}}_{n}$ is discrete, only the path connected components of $\Om^n S^k$ matter in Eq.~\eqref{loopSphereOP}, so
\begin{equation}\label{DiscloopSphereOP}
    [\Om^n S^k,\mathcal{E}^{\txt{discrete}}_{n}]_{\EE_n} = [\pi_0(\Om^n S^k),\mathcal{E}^{\txt{discrete}}_{n}]_{\EE_n} = \mathrm{Hom}(\pi_n( S^k),\mathcal{E}^{\txt{discrete}}_{n}).
\end{equation}
Since both ${\pi_0(\Om^nS^k) = \pi_n( S^k)}$ and $\mathcal{E}^{\txt{discrete}}_{n}$ are discrete, the $\EE_n$-algebra maps are group homomorphisms. Notice that because ${\pi_{n}(S^k) = 0}$ when ${k>n}$ and ${\pi_k(S^k) = \Z}$, Eq.~\eqref{DiscloopSphereOP} agrees with~\eqref{loopSphereOP3} for ${\mathcal{E}_n = \mathcal{E}_n^{\txt{discrete}}}$. Interestingly,~\eqref{DiscloopSphereOP} is generally nontrivial when ${k < n}$ as $\pi_n(S^k)$ is generally nontrivial. For example, when ${n=3}$ and ${k=2}$,
\begin{equation}
    [\Om^3 S^2,\mathcal{E}^{\txt{discrete}}_{3}]_{\EE_3} = \mathcal{E}^{\txt{discrete}}_{3},
\end{equation}
because ${\pi_3(S^2) = \Z}$. Since the $\pi_1(B^n\mathcal{E}_n)$ action is trivial for $n>1$, $3$-brane fields with discrete $\mathcal{E}_3$ have codimension $3$ topological solitons classified by $\mathcal{E}_3$.

Let us now consider the example $\mathcal{E}_n = U(1)$. Since $U(1)$ is an abelian group, we can view it as a grouplike $\EE_n$-algebra for any $0 \leq n \leq \infty$. For $\mathcal{E}_n = U(1)$, we find\footnote{We refer the interested reader to Appendix~\ref{U(1)HomotopyTheoryForFunAppendix} for the calculation of ${[\Om^n S^k,U(1)]_{\EE_n}}$.}
\begin{equation}\label{EnU1}
     [\Omega^n S^k, U(1)]_{\EE_n} =  
     \begin{cases}
             0 \quad \quad & n < k-1,\\
        \mathbb{Z} \quad\quad &n = k-1, \\ 
        0 \quad \quad & n = k = 2\\
         \pi_{n}(S^k)\oplus \mathbb{Z} \quad\quad &n = 2k -2,~ k \in 2 \mathbb{Z},~ k > 2 \\ 
          \pi_{n}(S^k)^{\txt{tor}} \quad\quad &\txt{else},
     \end{cases}
\end{equation}
where ${\pi_{n}(S^k)^{\txt{tor}}}$ is the torsion part of $\pi_{n}(S^k)$ and $\oplus$ is direct sum.\footnote{Mathematically, it is more natural to replace ${\pi_{n}(S^k)^{\txt{tor}}}$ with its Pontryagin dual ${\mathrm{Hom}(\pi_{n}(S^k)^{\txt{tor}}, U(1))}$. However, they are abstractly isomorphic.} Notice that this reproduces Eq.~\eqref{loopSphereOP3} when ${k\geq n}$. Setting ${k = 2}$, for example, this becomes
\begin{equation}
    [\Om^n S^2,U(1)]_{\EE_n} \simeq 
         \begin{cases}
        0 \quad \quad & n = 0,\\
        \mathbb{Z} \quad\quad &n = 1, \\ 
         0 \quad\quad &n = 2, \\ 
         0 \quad \quad &n=3, \\ 
        \pi_{n}(S^2) \quad\quad & n > 3,
     \end{cases}
\end{equation}
When ${n=1}$, the topological solitons correspond to the codimension 3 magnetic monopoles in $1$-form $U(1)$ gauge theory. The others are beyond gauge fields. Interestingly, $\pi_3(S^2)$ generated by the Hopf map $S^3 \to S^2$ does not contribute since it is torsion-free. On the other hand, since ${\pi_4(S^2) = \mathbb{Z}_2}$ is torsion, we have
\begin{equation}
    [\Omega^4 S^2, U(1)]_{\mathbb{E}_4} = \mathbb{Z}_2.
\end{equation}
This implies that $4$-brane fields with $\mathcal{E}_4 = U(1)$ have codimension 3 topological solitons classified by $\Z_2$.

\section{Application to spontaneous symmetry breaking}\label{applicationToSSBSec}

Since an $n$-form symmetry's charge is carried by $n$-dimensional objects in space, $n$-brane fields naturally play the role of order parameters for diagnosing spontaneously broken $n$-form symmetries. In this section, we discuss the topological solitons arising from spontaneously breaking an invertible $n$-form symmetry described by the group $G^{(n)}$. We will consider only ${n>0}$, which makes $G^{(n)}$ an abelian group~\cite{GW14125148}, and discuss the topological solitons using the $n$-brane field framework.

\subsection{Solitons of higher-form symmetry breaking}

Consider the spontaneous symmetry breaking (SSB) pattern
\begin{equation}\label{nFormSSBpattern}
    G^{(n)}\ssb H^{(n)},
\end{equation}
where ${H^{(n)}\subset G^{(n)}}$. This SSB signals a phase where the ground states (i.e., vacua) are invariant under only the subgroup $ H^{(n)}$ of $G^{(n)}$ and form an orbit under ${G^{(n)}/H^{(n)}}$. A phase with this SSB pattern is characterized by an $n$-dimensional order parameter carrying a $G^{(n)}$ symmetry charge corresponding to~\eqref{nFormSSBpattern}. In particular, it is the $n$-brane field
\begin{equation}
    \Psi\colon \txt{Brane}_n(X_D)\to G^{(n)}/H^{(n)},
\end{equation}
where the target space is the coset space
\begin{equation}
    G^{(n)}/H^{(n)} = \{g H^{(n)}~|~g\in G^{(n)} \}.
\end{equation}
However, since $G^{(n)}$ is abelian, $H^{(n)}$ is always a normal subgroup and $G^{(n)}/H^{(n)}$ is a group. Therefore, we can view $G^{(n)}/H^{(n)}$ as a grouplike $\EE_n$-algebra and consider the homomorphism $n$-brane field
\begin{equation}
    \psi\colon \Om^n X_D\to \mathcal{E}_n = G^{(n)}/H^{(n)}.
\end{equation}
In what follows, we apply the classification developed in Sec.~\ref{nBraneDefects} to the $n$-brane $\psi$'s topological solitons.

Let us first consider the case where ${n=1}$ or $2$, where the codimension $(k+1)$ topological solitons of $\psi$ are classified by
\begin{equation}\label{eq:n=1-helper}
    \pi_k(B^n \mathcal{E}_n)/\pi_1(B^n \mathcal{E}_n) = \pi_{k-n}(\mathcal{E}_n)/\pi_{1-n}(\mathcal{E}_n).
\end{equation}
for all $k$. This matches the classification of $n$-form $ \mathcal{E}_n$ gauge theory. Let us quickly discuss the $\pi_1$ action. The $\pi_1$ action is always trivial when ${n>1}$. However, because $\mathcal{E}_n$ is an abelian group, it is also trivial when ${n=1}$. Indeed, when ${n=1}$, the ${\pi_1(B\mathcal{E}_1)}$ action on ${\pi_k(B\mathcal{E}_1)}$ becomes a ${\pi_0(\mathcal{E}_1)}$ action on ${\pi_{k-1}(\mathcal{E}_1)}$ given the following: given ${a \in \mathcal{E}_1}$ representing ${[a] \in \pi_0(\mathcal{E}_1}$ and ${\gamma^{(k-1)}\colon S^{k-1} \to \mathcal{E}_1}$ representing ${[\gamma^{(k-1)}] \in \pi_{k-1}(\mathcal{E}_1)}$, then ${[a] \cdot [\gamma^{(k-1)}]}$  in ${\pi_{k-1}(\mathcal{E}_1)}$ is represented by 
\begin{equation}
    a  \gamma^{(k-1)} a^{-1} \colon S^{k-1} \xrightarrow{ \gamma^{(k-1)}}\mathcal{E}_1 \xrightarrow{a(-)a^{-1}}\mathcal{E}_1,
\end{equation}
where the latter is the conjugation action by $a$. Since $\mathcal{E}_1$ is an abelian group, $a(-)a^{-1}$ is equal to identity and, therefore, the $\pi_0(\mathcal{E}_1)$ action is trivial. 

Therefore, when ${n=1}$ or $2$, the codimension $(k+1)$ topological solitons of $\psi$ are classified by ${\pi_{k-n}(\mathcal{E}_n)}$. For instance, if $G^{(n)} = \Z_N$ and ${H^{(n)} = \Z_r}$, where $r$ is any divisor of $N$, they are classified by
\begin{equation}
    \pi_k(B^n[\Z_N/\Z_r]) = \begin{cases}
        \Z_{N/r}\quad\quad & k = n,\\
        0\quad\quad & k \neq n.\\
    \end{cases}
\end{equation}
This SSB phase corresponds to the deconfined phase of $\Z_{N/r}$ $n$-form gauge theory, and the topological solitons are the $\Z_{N/r}$ codimension $n+1$ magnetic defects. Let us now consider $G^{(n)} = U(1)$ and $H^{(n)} = \Z_N$, corresponding to a Coulomb phase where deconfined gauge charges carry charge $N\Z$. The topological solitons are classified by
\begin{equation}
    \pi_k(B^n[U(1)/\Z_N]) = \begin{cases}
        \Z\quad\quad & k = n+1,\\
        0\quad\quad & k \neq n+1,\\
    \end{cases}
\end{equation}
which is the codimension $n+2$ magnetic defect in the Coulomb phase.

We now discuss the case where ${n>2}$. When ${k\geq n}$, using Eq.~\eqref{loopSphereOP3}, codimension $k+1$ topological solitons are classified by
\begin{equation*}
    \pi_{k-n}(\mathcal{E}_n) = \pi_k(B^n G^{(n)}/B^n H^{(n)}).
\end{equation*}
The examples for $G^{(n)} = \Z_N$ and $U(1)$ given above generalize straightforwardly. Furthermore, this result agrees with a conjecture made in Ref.~\citenum{S150804770} stating that these solitons are classified by the homotopy groups of ${B^n G^{(n)}/B^n H^{(n)}}$. For $k<n$, the topological solitons of $\psi$ no longer agree with those arising from $n$-form gauge fields: when $G^{(n)}/H^{(n)}  = \Z_N/\Z_r = \Z_{N/r}$, since $\Z_{N/r}$ is discrete, we can apply Eq.~\eqref{DiscloopSphereOP} and get that codimension $k+1$ defects are classified 
\begin{equation}
    \mathrm{Hom}(\pi_n( S^k),\Z_{N/r}).
\end{equation}
On the other hand, when $G^{(n)}/H^{(n)} = U(1)/\Z_N \simeq U(1)$, Eq.~\eqref{EnU1} gives us the classification of solitons. Since $n$-form symmetries are typically discussed in the context of $n$-form gauge fields (hence its name), it seems unnatural to attribute these topological solitons as arising from spontaneously broken $n$-form symmetries. Perhaps there is a more general notion of $n$-brane symmetry whose spontaneous breaking is described by generic homomorphism $n$-brane fields (including those beyond gauge theory). We leave this interesting possibility for future investigations.

\subsection{Comments on 't Hooft anomalies}

Having discussed when the topological solitons of the homomorphism $n$-brane fields can be understood as arising from spontaneously broken $n$-form symmetry, here we make a brief remark about topological solitons of spontaneously broken $n$-form symmetries. In particular, we note an interesting observation that uniquely affects ${n>0}$. Namely, topological solitons arising from anomalous higher-form symmetries must be non-dynamical; otherwise, they explicitly break parts of, or all of, the spontaneously broken symmetry. We will demonstrate this in three well-known examples: Maxwell theory, BF theory, and $U(1)$ Chern-Simons theory.

\subsubsection{Maxwell Theory}

Let us first consider Maxwell theory, which we consider in ${D=4}$ dimensional spacetime for concreteness and constructed from the $1$-form $U(1)$ gauge field $a$. Its action is
\begin{equation}
    S = \frac{1}{2e^2}\int_{X_4} \dd a\wdg\hstar \dd a
\end{equation}
and has a spontaneously broken anomalous $U_e(1)\times U_m(1)$ $1$-form symmetry generated by
\begin{equation}
    D_2^{(\al_e,\al_m)}(\Si) = \exp[\ii\int_\Si \left(\frac{\al_e}{e^2}\hstar \dd a + \frac{\al_m}{2\pi}\dd a\right)],
\end{equation}
where $\Si$ is a closed 2-submanifold in spacetime $X_4$ and $\al_e$ and $\al_m$ parameterize the electric and magnetic symmetries $U_e(1)$ and $U_m(1)$, respectively. Since the SSB pattern corresponds to ${G^{(1)}\ssb H^{(1)}}$ with ${G^{(1)} = U_e(1)\times U_m(1)}$ and ${H^{(1)} = 1}$, the resulting solitons are classified by ${\Z\times \Z}$. Parametrizing this group by ${(q_e,q_m)\in \Z\times \Z}$, a general soliton corresponds to the dyon line 
\begin{equation}
    D^{(q_e,q_m)}(\ga) = W^{q_e}(\ga) T^{q_m}(\ga),
\end{equation}
where $\ga$ is a loop in $X_4$ and  $W(\ga)$ and $T(\ga)$ are electric (Wilson) and magnetic ('t Hooft) defect lines arising from $U_m(1)$ and $U_e(1)$ spontaneously breaking, respectively. Indeed, $U_m(1)$ is generated by the winding number of the photon $a$ while $U_e(1)$ is generated by the winding number of the dual photon. Notice how the soliton of the electric (magnetic) symmetry is the charged object of the magnetic (electric) symmetry. This is a consequence of the 't Hooft anomaly.

This dyon line is not dynamical, describing the worldline of an infinitely massive dyon. It can be made dynamical by introducing electric and magnetic matter fields. However, this explicitly breaks the ${U_e(1)\times U_m(1)}$ $1$-form symmetry. In particular, making the soliton of $U_e(1)$ dynamical explicitly breaks $U_m(1)$ and vice versa.

\subsubsection{BF Theory}

We next consider $\Z_N$ BF theory in ${D=4}$. Introducing the $U(1)$ $1$-form gauge field $a$ and $U(1)$ $2$-form gauge field $b$, its action is given by
\begin{equation}
    S = \frac{N}{2\pi}\int_{X_4} a \wdg \dd b.
\end{equation}
The equations of motion make $a$ and $b$ flat and quantize their holonomies as
\begin{equation}
    \int_\ga a,~ \int_\Si b \in \frac{2\pi}{N} \Z.
\end{equation}
It is well known that the theory has a spontaneously broken anomalous $\Z_N^{(1)}\times \Z_N^{(2)}$ symmetry, where $\Z_N^{(1)}$ denotes the electric $\Z_N$ $1$-form symmetry and $\Z_N^{(2)}$ the magnetic $\Z_N$ $2$-form symmetry. It is generated by the symmetry defects
\begin{equation}\label{BFsymDefects}
    D_2^{(\al_e)}(\Si) = \exp[\ii \al_e\int_\Si b]\quad\quad\quad D_1^{(\al_m)}(\ga) = \exp[\ii \al_m \int_\Si a],
\end{equation}
where $\al_e,\al_m\in\{0,1,\cdots,N-1\}$ parameterize the symmetry group. The SSB pattern is described by ${(G^{(1)},G^{(2)})\ssb (H^{(1)},H^{(2)})}$ where ${G^{(1)} = G^{(2)} = \Z_N}$ and ${H^{(1) }= H^{(2)} = 1}$. The solitons arising from this SSB pattern are codimension 2 and 3, classified by $\Z_N$, and because the symmetry is discrete, they are the symmetry defects~\eqref{BFsymDefects}. Again, as a consequence of the anomaly, the soliton arising from spontaneously breaking one symmetry carries the symmetry charge of the other symmetry.

The solitons~\eqref{BFsymDefects} are not dynamical. They can be made dynamical by introducing new degrees of freedom. For instance, $D_2^{(\al_e)}$ can be made dynamical by introducing a $1$-form field that minimally couples to $b$ and $D_1^{(\al_m)}$ can be by introducing electric matter of $a$. However, doing so will explicitly break the symmetry. For example, making the defect $D_2^{(\al_e)}$, which arises from breaking $\Z_N^{(1)}$, dynamical explicitly breaks $\Z_N^{(2)}$. We again see that due to the mixed anomaly, making the solitons arising from spontaneously breaking higher-form symmetries dynamical explicitly breaks said higher-form symmetries.

\subsubsection{\texorpdfstring{$U(1)_k$}{U(1)k} Chern-Simons Theory}

Lastly, we next consider $U(1)$ level $k\in \Z$ Chern-Simons theory. Letting $a$ denote a $U(1)$ 1-form gauge field and $X_3$ ${D=3}$ dimensional spacetime, it is described by the action
\begin{equation}
    S = \frac{k}{4\pi}\int_X a\wdg\dd a.
\end{equation}
It is well known that this theory has a spontaneously broken anomalous $\mathbb{Z}_k$ 1-form symmetry generated by
\begin{equation}\label{CSsymDef}
    D_1^{(\al)}(\ga) = \exp[\ii \frac{\al k}{2\pi} \int_\ga a],
\end{equation}
where ${\al\in\{0,\frac{2\pi}k,\cdots, \frac{2\pi(k-1)}k\}}$ and $\ga$ is a loop in spacetime $X_3$. Since the SSB pattern is ${G^{(1)}\ssb H^{(1)}}$ with ${G^{(1)} = \Z_k}$ and ${H^{(1)} = 1}$, there is a one-dimensional soliton classified by $\Z_k$ and it is the symmetry defect~\eqref{CSsymDef}. Once again, a manifestation of the 't Hooft anomaly is that this soliton carries the $\Z_k$ 1-form symmetry charge.

Again, the defect~\eqref{CSsymDef} is not dynamical. It can be made dynamical by adding electric matter to the theory, but doing so explicitly breaks the $\Z_k$ 1-form symmetry.

\section{Outlook}\label{outlook}

In this note, we investigated the topological aspects of brane fields by classifying topological solitons of homomorphism $n$-brane fields defined in Sec.~\ref{meanBraneField}. Our main result, which we presented in Sec.~\ref{sec:ClassDefs}, was classifying these topological solitons using maps of $\EE_n$-algebras that arise from the $n$-brane fields. We gave many examples and contextualized them to higher-form symmetry breaking phases in Sec~\ref{applicationToSSBSec}.

One of our motivations for considering brane fields was their utility in providing order parameters for higher-form symmetries. It is natural to wonder what the order parameters of other generalized symmetries are. For example, it would be interesting to investigate order parameters of higher group symmetries~\cite{KT13094721, S150804770, CI180204790, BH180309336, BCH221111764}. For example, invertible $0$-form and $1$-form symmetries can mix nontrivially and form a mathematical structure called a $2$-group. If this $2$-group splits, meaning that the $0$-form and $1$-form symmetries do not mix, then the solitons arising from its spontaneous breaking are simply those of the $0$-form and $1$-form symmetries independently. However, for nontrivial $2$-groups, the solitons of one symmetry will generally influence the solitons of the others. It would be interesting to formulate local order parameter fields that interact with string order parameter fields nontrivial to mimic the $2$-group's symmetry charges and investigate the resulting topological solitons.

\section*{Acknowledgements}

We are grateful for John Mcgreevy for comments on the manuscripts as well Atakan Hilmi Fırat, Hart Goldman, Cameron Krulewski, John McGreevy, Daniel Ranard, and Xiao-Gang Wen for helpful discussions. S.D.P. is supported by the National Science Foundation Graduate Research Fellowship under Grant No. 2141064 and by the Henry W. Kendall Fellowship.
Y.L.L. is supported by the Simons Collaboration on Global Categorical Symmetries.

\begin{appendix}

\section{Based and free homotopy classes}\label{basedVsFreeAppen}

In this appendix, we will review the difference between free and based homotopy classes. We refer the reader to Sec. 1.4 of Ref.~\citenum{MayPonto12} for a more rigorous discussion.

Given two spaces $A$ and $B$, a \textit{free}\footnote{as opposed to based} map is simply a map from $A$ to $B$.  Furthermore, given two maps ${f, g\colon A \to B}$,  a \textit{free homotopy} from $f$ to $g$ is a map ${h\colon A \times I \to B}$ such that ${h(-, 0) = f}$ and ${h(-, 1) = g}$. Letting $t$ parameterize $I$, we can view ${h_t(-) = h(-, t)}$ as a family of maps from ${A}$ to ${B}$, continuously deforming from $f$ to $g$. We denote ${[A, B]_\txt{f}}$ the set of homotopy equivalent classes of free maps from $A$ to $B$. We denote the free homotopy class of a free map $f$ as ${[f]_{\txt{f}} \in [A ,B]_\txt{f}}$. Therefore, if two free maps $f$ and $g$ are connected by a homotopy, then ${[f]_\txt{f} = [g]_\txt{f}}$, and otherwise ${[f]_\txt{f} \neq [g]_\txt{f}}$.

Perhaps more familiar is the variant of based maps and based homotopies. Let us assume the spaces $A$ and $B$ have basepoints $a$ and $b$, respectively, which we signify by denoting them as ${(A,a)}$ and ${(B,b)}$. A \textit{based} map ${f \colon (A,a) \to (B,b)}$ is a map from $A$ to $B$ such that ${f(a) = b}$. Given two based maps ${f, g\colon (A,a) \to (B,b)}$, a \textit{based homotopy} from $f$ to $g$ is a free homotopy ${h \colon A \times I \to B}$ such that ${h(a, t) = b}$ for all ${t \in I}$. That is, it is a continuous deformation from $f$ to $g$ through based maps. We denote the set of based homotopy classes of based maps from $A$ to $B$ as ${[(A, a), (B, b)]_\txt{b}}$. We denote the based homotopy class of a based map $f$ as ${[f]_{\txt{b}} \in [(A, a), (B, b)]_\txt{b}}$.

These definitions demonstrate the differences between the free homotopy classes ${[A, B]_\txt{f}}$ and based homotopy classes ${[(A, a),(B,b)]_\txt{b}}$ of maps from $A$ to $B$. We now discuss how, despite these differences, the two are related. Indeed, since based maps form a subset of free maps, and based homotopies form a subset of free homotopies, there exists a map of sets 
\begin{equation}\label{eq:based-to-free}
    [(A, a), (B, b)]_{\txt{b}} \to [A, B]_{\txt{f}}.
\end{equation}

\begin{figure}[t!]
    \centering
        \includegraphics[width=.91\textwidth]{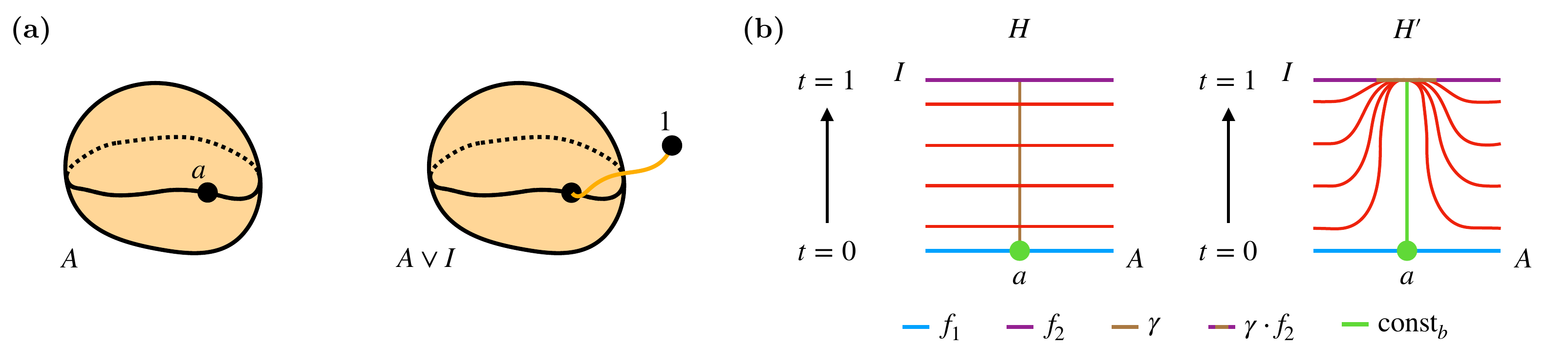}
        \caption{(a) ${A \vee I}$ is constructed from $A$ by gluing ${0\in I}$ to ${a\in A}$. It is a based space with  ${1\in I}$ being its basepoint.
        (b) We construct a based homotopy $H'$ between ${f_1}$ and ${\ga\cdot f_2}$ using a free homotopy $H$ between $f_1$ and $f_2$. $H$ is shown on the left, in the case where $A$ is an interval and $f_1$ and $f_2$ colored in blue and purple, respectively. The loop $\ga$ is traced out by the basepoint $a$ under $H$. The red lines depict horizontal slices of $H$ at fixed values of ${t\in I}$. The based homotopy $H'$ is shown on the right. It is constructed by ``pushing'' all of $\gamma$ in $H$ to ${t = 1}$, which also lifts the horizontal red slices as shown. Therefore, ${H'(a, t) = b}$ for all ${t \in I}$, which we denote by the green line.}
        \label{fig:balloon}
\end{figure}

To understand~\eqref{eq:based-to-free}, we must first discuss how moving the basepoint affects based homotopy classes. Consider a based map ${f\colon (A, a) \to (B, b)}$ and a path $\delta$ from the basepoint ${b\in B}$ to another point ${b'\in B}$. $\delta$ can be thought of as the path taken when moving the basepoint of $B$ from $b$ to $b'$. 
Using $f$ and $\delta$, we can construct a based map from ${(A, a)}$ to ${(B, b')}$. First, we consider the space ${A \vee I}$, which can be constructed by gluing ${0\in I}$ to the basepoint ${a\in A}$, and whose basepoint we set to ${1\in I}$ (see Fig.~\ref{fig:balloon}a). 
Then, we consider the based map ${f \vee \delta \colon (A \vee I,1) \to (B,b')}$, which restricts to $f$ on $A$ and $\del$ on $I$. Since ${A \vee I}$ is homotopic equivalent to $A$ by shrinking the interval $I$, $f \vee \del$ induces a based map $\del \cdot f: (A, a) \to (B, b')$. Not only does this construction yield a based map from ${(A, a)}$ to ${(B, b')}$, as claimed, it also defines an action of paths on based homotopy classes:
\begin{equation}\label{pathAction}
      \del \cdot - \colon [(A, a), (B, b)]_{\txt{b}} \to  [(A, a), (B, b')]_{\txt{b}}.
\end{equation}
Furthermore, shrinking the interval $I$ gives a free homotopy between $f$ and $\del \cdot f$, thus  ${[f]_\txt{f} = [\del\cdot f]_\txt{f}}$ in $[A, B]_{\txt{f}}$.

In the case that ${b = b'}$, $\del$ becomes a loop $\gamma$ based at $b$ and ${\ga \cdot f}$ is another based map from $(A, a)$ to $(B, b)$. Eq.~\eqref{pathAction} then defines an action of based loops in $(B,b)$ on $[(A, a), (B, b)]_{\txt{b}}$. In fact, given two based loops $\gamma, \gamma'$  in $(B,b)$, if $[\gamma]_{\txt{b}} = [\gamma']_{\txt{b}}$ in $\pi_1(B,b)$ then $[\gamma \cdot f]_{\txt{b}} = [\gamma' \cdot f]_{\txt{b}}$. Therefore Eq.~\eqref{pathAction} descents to an action of $\pi_1(B, b)$ on $[(A, a), (B, b)]_{\txt{b}}$. Notice that $[f]_{\txt{f}} = [\gamma \cdot f]_{\txt{f}}$ even though $[\gamma \cdot f]_{\txt{b}}$ might not equal $[f]_{\txt{b}}$, making $\ga\cdot -$ a map that relates freely homotopic maps that are not based homotopic. In particular, Eq.~\ref{eq:based-to-free}
descents to a map 
\begin{equation}\label{eq:based-to-free2}
    [(A, a), (B, b)]_{\txt{b}}/\pi_1(B,b) \to [A, B]_{\txt{f}}.
\end{equation}

We now show that when $B$ is path-connected (i.e., any two points in $B$ are connected by a path), Eq~\eqref{eq:based-to-free2} is a bijection:
\begin{equation}
    [(A, a), (B, b)]_{\txt{b}}/\pi_1(B, b) \simeq [A, B]_\txt{f}.
\end{equation}
We prove this by showing it is surjective and injective.
\begin{enumerate}
    \item For surjectivity, we need to show that any free map ${f' \colon A \to B}$ is free homotopy equivalent to a based map from ${(A, a)}$ to ${(B, b)}$. To show this, letting ${b' = f'(a)}$, we can first view $f'$ as a based map from ${(A, a)}$ to  ${(B, b')}$. Since $B$ is path-connected, there is a path $\del$ from $b'$ to $b$, and by the above construction, we have a based map ${\del\cdot f'\colon (A ,a)\to (B,b)}$ for which ${[f']_\txt{f} = [\del\cdot f']_\txt{f}}$, making Eq.~\eqref{eq:based-to-free2} surjective.
    \item For injectivity, given two based maps $f_1, f_2 \colon {(A, a)} \to {(B, b)}$, if ${[f_1]_\txt{f} = [f_2]_\txt{f}}$, then we need to show that there exists a based loop $\gamma$ at $(B, b)$, such that $[f_1]_{\txt{b}} = [\gamma \cdot f_2]_{\txt{b}}$. Let $H$ be a \textit{free} homotopy between $f_1$ and $f_2$. Since $f_1$ and $f_2$ are based maps, ${\gamma(t) \coloneqq H(a, t)}$ is a based loop in $(B,b)$ since ${H(a, 0) = H(a,1) = b}$. Then, using $H$, we can now construct a \textit{based} homotopy $H'$ between $f_1$ to $\gamma \cdot f_2$, as shown in Fig.~\ref{fig:balloon}b. Thus, Eq.~\eqref{eq:based-to-free2} is injective.
\end{enumerate}

\section{Introduction to iterated loop spaces}\label{LoopSpaceAppendix}

In this appendix section, we review iterated loop spaces, the algebraic structure they follow called grouplike $\EE_n$-algebras, and delooping constructions. Assuming familiarity with basic homotopy theory (i.e., homotopy groups), our goal is to provide a pedagogical, non-rigorous introduction for physicists. We refer the reader to Ref.~\citenum{May06} for a more comprehensive introduction.

\subsection{Iterative loop spaces}\label{itLoopSpaceIntro}

\begin{figure}[t!]
    \centering
        \includegraphics[width=.91\textwidth]{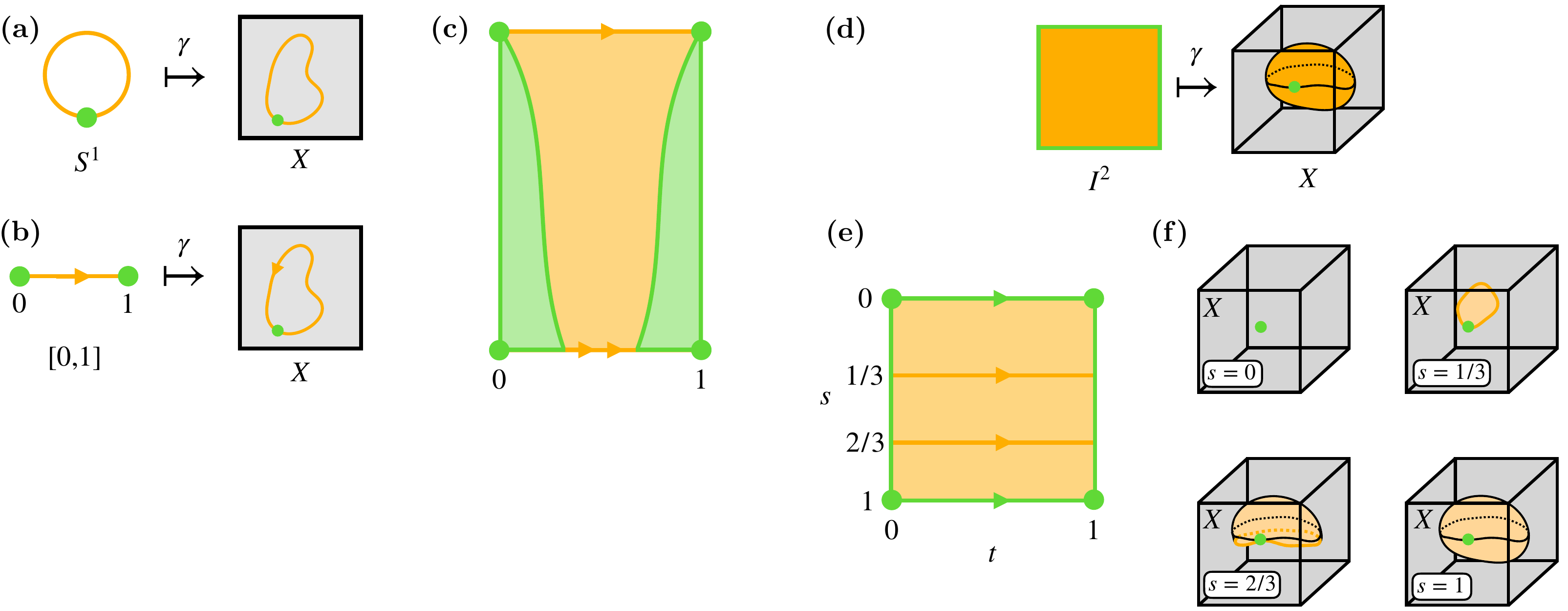}
        \caption{(a) A loop in a space $X$ defines a map $\ga$ from $S^1$ to $X$, which maps the basepoint of $S^1$ to the basepoint of $X$ (which are both colored in green). (b) Equivalently, it is a map from the interval ${[0,1]}$ to $X$ that takes the boundary ${\{0,1\}}$ to the basepoint. (c) The path in (b), shown on top, can be deformed to the loop shown on the bottom part of the interval, not just the boundary ${\{0,1\}}$ is mapped to the basepoint of $X$. As one moves downward in the rectangle shown, each horizontal slice corresponds to a loop that stays at the basepoint longer but moves faster around the loop. (d) A $2$-fold loop ${\ga\colon  S^2 \to X}$ is equivalent to a map ${I^2 \to X}$ such that the boundary $\partial I^2$ maps to the basepoint of $X$ (both shown in green). (e) Parameterizing $I^2$ by ${(s,t)}$, given a $2$-fold loop $\gamma\colon  I^2 \to X$, for any fixed $s$, $\gamma(s)\colon  I \to X$ is a loop in $X$. Therefore, as $s$ varies from $0$ to $1$, we form a loop of loops in $X$. (f) Here, this process is shown at $s$-slices $s=0$, $1/3$, $2/3$, and $1$, with the loop in $X$ at each $s$-slice draw in dark orange while the surface it traced out in $X$ formed by the loops of loops shaded in light orange. Starting from the constant loop (shown in green) at ${s=0}$, the loop grows and moves to form the northern hemisphere of the 2-fold loop, wraps around the back to form the southern hemisphere, and then shrinks to return to the constant loop at ${s=1}$.} 
        \label{fig:loops}
\end{figure}

Consider a space $X$ with a basepoint ${x\in X}$. Given $X$, one can construct a new space made up of all possible loops in $X$, which is called the loop space\footnote{This is also referred to as the based loop space, as opposed to the free loop space, which considers loops in $X$ without the constraint that the basepoint is mapped to a specified point $x \in X$.} of $X$ and denoted as $\Om X$. A point in $\Om X$ is a loop ${\ga\colon  S^1 \to X}$ that takes the basepoint of $S^1$ to the basepoint ${x\in X}$, as demonstrated in Fig.~\ref{fig:loops}a. However, it is convenient to instead view a point in $\Om X$ as a path ${\gamma\colon  I \to X}$, where ${I\equiv[0,1]}$, that starts and ends at the basepoint $x$, as shown in Fig.~\ref{fig:loops}b.

Let us note two additional important properties of loop spaces, which we will use later. Firstly, there is the so-called constant loop, which is the loop formed from mapping the entire path $I$ to the basepoint of $X$. The constant loop at $X$ is a natural basepoint for the loop space $\Om X$. Furthermore, there is an intrinsic notion of two loops being close to one another in $\Om X$, as they can be deformed into one another. This makes $\Om X$ a topological space. 

Instead of considering the space of all possible loops in $X$, one can generalize the above discussion and consider the space of all possible ``$n$-fold loops'' in $X$. Here, $n$-fold loop is an $n$-dimensional loop in $X$, i.e., a map $S^n \to X$ that takes the basepoint of $S^n$ to $x$. The $n$-fold loop space of $X$ is denoted as $\Om^n X$.  
However, as shown for the ${n=2}$ case in Fig.~\ref{fig:loops}d, this map is equivalent to ${I^n \to X}$ with the entire boundary $\partial I^n$ mapping to the basepoint of $X$. Lastly, we note that like $\Om X$, $\Om^n X$ is a topological space with a canonical basepoint (the constant $n$-fold loop).

It turns out that the $n$-fold loop space of $X$ is related to the ${m<n}$ loop space of $X$. Let us motivate this relationship explicitly in the ${n=2}$ case, after which the generalization to arbitrary $n$ will be straightforward.

Consider a 2-fold loop ${\Ga\in\Om^2 X}$ as the map $\Gamma\colon  I^2 \to X$ discussed above, where $\pp I^2$ maps to the basepoint ${x\in X}$. We can parameterize $\Ga$ by introducing coordinates ${(s, t)}$ of $I^2$ to construct the function ${\Ga(s,t)}$. For each fixed $s$-slice ${s = s_0}$, ${\Gamma(s_0, t)}$ defines a path in $X$ as $t$ is varied from $0$ to $1$, which we denote as $\Gamma_{s_0}(t)$. Since the points ${(s_0, 0)}$ and ${(s_0,1)}$ are on the boundary of $I^2$, they must map to the basepoint $x$. Consequently, for each $s$-slice $s_0$, $\Gamma_{s_0}$ is a loop in $X$. Consider the $s$-slices ${s = 0}$ and ${s = 1}$ which reside on the boundary $\pp I^2$. By definition, $\Ga_{s_0=0,1}(t)$ must be the basepoint $x$ for all $t$, and therefore $\Ga_{s_0=0}$ and $\Ga_{s_0=1}$ are both the constant loop (i.e., the basepoint of $\Omega X$). So, varying $s_0$ from $0$ to $1$, $\Gamma_{s_0}$ starts at the constant loop, explores parts of $\Om X$, and then returns back to the constant loop, thus forming a 1-fold loop in $\Om X$. However, since $\Ga_{s_0}$ itself is a 1-fold loop in $X$, $\Gamma$ is a 1-fold loop of 1-fold loops in $X$ and defines a point in $\Omega(\Omega X)$ (see Fig~\ref{fig:loops}e). Therefore, each 2-fold loop in $X$ corresponds to a 1-fold loop in $\Om X$ and so
\begin{equation}\label{2loopand1loops}
    \Om^2 X = \Om(\Om X).
\end{equation}

Let us now generalize the above and consider a ${(p+q)}$-fold loop ${\Ga\in\Om^{p+q}X}$. The coordinates of $I^{p+q}$ can always be labeled as ${(s_1,\cdots, s_p, t_1, \cdots t_q)}$. For a fixed ${(s_1,\cdots, s_p)}$-slice, ${\Ga_{s_1,\cdots s_p}(t_1,\cdots, t_q)}$ forms a $q$-fold loop in $X$ upon varying each $t_i$ since the boundary $\pp I^{q+p}$ must map to $x$. Then, varying each ${s_i}$ from 0 to 1, $\Ga_{s_1,\cdots s_p}$ forms a $p$-fold loop in $\Om^q X$. Therefore, in general, each ${(p+q)}$-fold loop in $X$ corresponds to a $p$-fold loop in $\Om^q X$, and so
\begin{equation}\label{eq:HigherLoopSpace}
    \Omega^{p+q} X = \Omega^p(\Omega^q(X)).
\end{equation}
Clearly, the ${p=q=1}$ case reproduces Eq,~\eqref{2loopand1loops}.

Because the $n$-th loop space has a topology and basepoint, one can study the homotopy groups of $\Om^n X$. Let us first explicitly consider the ${n=1}$ case, after which we will generalize the discussion to the general $n$ case. All homotopy groups we consider are implicitly taken at the basepoint $x$.

First, consider $\pi_0(\Omega X)$, which is the set of path-connected components of $\Omega X$. If two loops $\gamma_1$ and $\gamma_2$ are in the same component, then there is a path in $\Omega X$ connecting $\gamma_1$ and $\gamma_2$. Transporting from $\ga_1$ to $\ga_2$ along this path in $\Om X$ is equivalent to continuously deforming $\ga_1$ into $\ga_2$ in $X$. Therefore, the path-connected components of $\Om X$ correspond to the classes of homotopy equivalent loops in $X$. Consequently,
\begin{equation}
    \pi_0(\Omega X) = \pi_1(X).
\end{equation}
Next, consider $\pi_1(\Om X)$, the set of homotopy-equivalent 1-fold loops in the loop space of $X$. Using that a 1-fold loop in $\Om X$ corresponds to a 2-fold loop in $X$, if two 1-fold loops are homotopy-equivalent in $\Om X$, then the two corresponding 2-fold loops in $X$ are also homotopy-equivalent. As a result, the 1st homotopy class of $\Om X$ corresponds to the second homotopy class of $X$:
\begin{equation}
    \pi_1(\Omega X) = \pi_2(X).
\end{equation}
There is an obvious pattern emerging from these two explicit cases. Indeed, consider $\pi_k(\Omega X)$, the set of homotopy-equivalent $k$-fold loops in the loop space of $X$. Since a $k$-fold loop in $\Om X$ corresponds to an $(k+1)$-fold loop in $X$, the $k$-th homotopy class of $\Om X$ corresponds to the $(k+1)$-th homotopy class of $X$:
\begin{equation}\label{eq:nHomotopy1loop}
    \pi_k(\Omega X) = \pi_{k+1}(X).
\end{equation}

It is straightforward to generalize this result for  $\Om^n X$. Indeed, using Eq.~\eqref{eq:HigherLoopSpace} we can write ${\Om^n X = \Om(\Om^{n-1} X)}$. Then, from Eq.~\eqref{eq:nHomotopy1loop}, the $k$-th homotopy class of ${\Om^n X}$ can be written as
\begin{equation*}
    \pi_k(\Om^n X) = \pi_k(\Om(\Om^{n-1} X)) = \pi_{k+1}(\Om^{n-1} X).
\end{equation*}
Repeating these manipulations $(n-1)$ more times, we find
\begin{equation}\label{nLoopkHomotopyraising}
    \pi_k(\Omega^n X) = \pi_{k+n}(X).
\end{equation}

\subsection{Loop concatenation and \texorpdfstring{$\EE_n$}{En} algebras}\label{appendixEn}

Given two $n$-fold loops ${\ga_1,\ga_2\in \Om^n X}$, one can fuse them together to form a new $n$-fold loop ${\ga_3\equiv \ga_1\circ\ga_2\in\Om^n X}$. This process, called loop concatenation, equips the $n$-fold loop space of $X$ with an algebra structure and makes $\Om^n X$ what is called a grouplike $\EE_n$-algebra. 

\begin{figure}[t!]
    \centering
        \includegraphics[width=.91\textwidth]{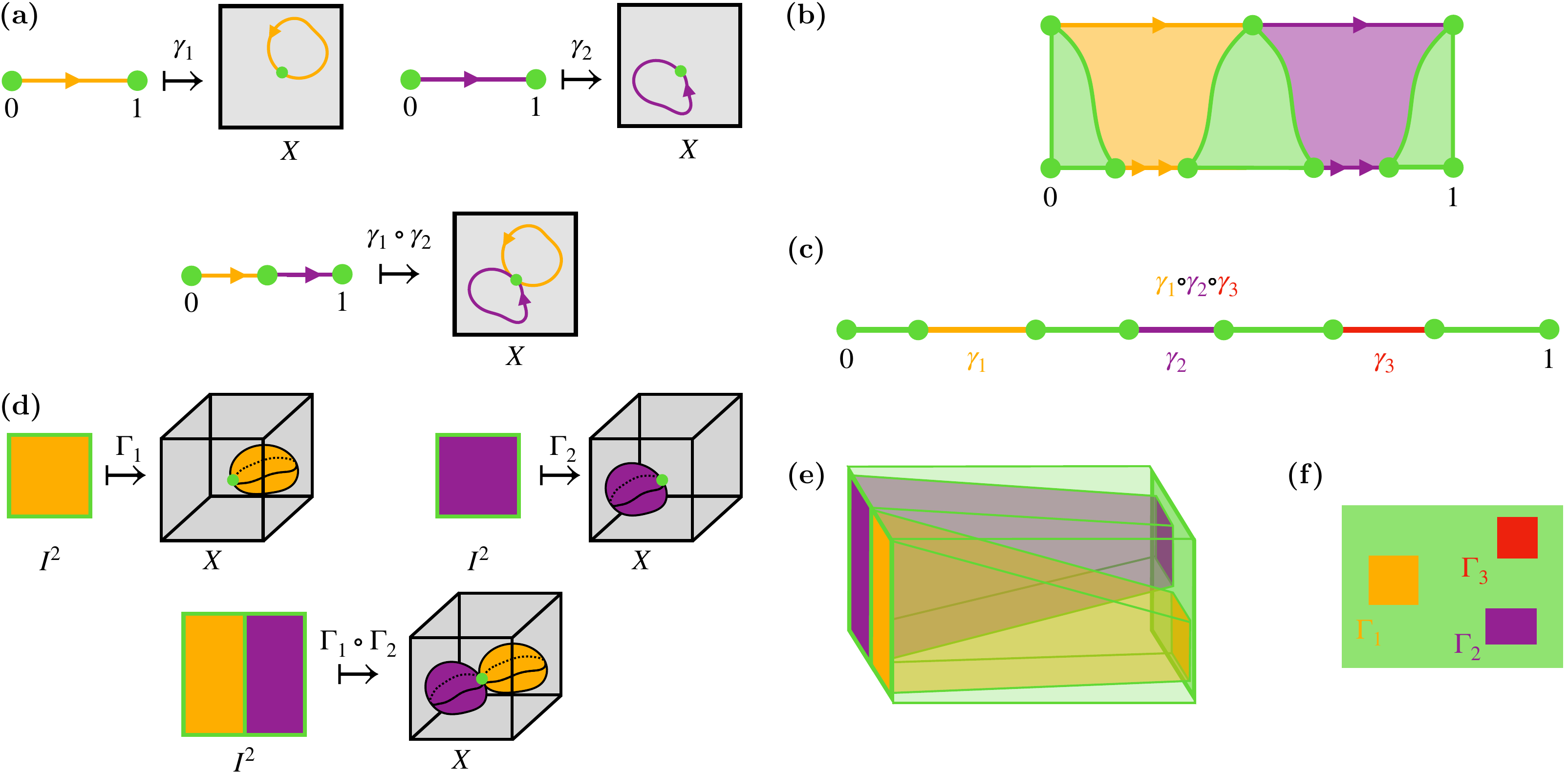}
        \caption{(a) Concatenating two 1-fold loops $\ga_1,\ga_2\colon I\to X$ produces a new 1-fold loop $\ga_1\circ\ga_2\colon I\to X$ formed by fusing $\ga_1$ and $\ga_2$ at the basepoint $x$ (shown in green). (b) However, since the interval $I$ matters up to homotopy (see Fig.~\ref{fig:loops}c), $\ga_1$ and $\ga_2$ in $\gamma_1 \circ \gamma_2$ do not need to be immediately concatenated. Instead, $\gamma_1$ and $\gamma_2$ in $I$ of $\ga_1\circ\ga_2$ can be deformed into line segments inside of a larger line such that $\ga_1\circ\ga_2$ maps all green line segments to the basepoint $x$. As long as the $\gamma_1$ line segment remains to the left or $\gamma_2$, the map from this deformed $I$ is homotopically equivalent to $\gamma_1 \circ \gamma_2$. (c) Using this, loop concatenation can be performed by embedding line segments labeled by 1-fold loops into the interval $I$. The relative distance and length of the line segments do not matter, but their ordering from left to right in $I$ does matter. (d) Concatenating two 2-fold loops $\Ga_1,\Ga_2\colon I\to X$ produces a new 2-fold loop $\Ga_1\circ\Ga_2\colon I\to X$ formed by fusing $\Ga_1$ and $\Ga_2$ at the basepoint $x$ (shown in green). (e) However, since the interval $I^2$ matters up to homotopy, $\Ga_1$ and $\Ga_2$ in ${\Gamma_1 \circ \Gamma_2}$ do not need to be immediately concatenated. Instead, $\Gamma_1$ and $\Gamma_2$ in $I^2$ of $\Ga_1\circ\Ga_2$ can be deformed into small squares inside of $I^2$ such that $\Ga_1\circ\Ga_2$ maps all green area to the basepoint $x$. (f) Using this, 2-fold loop concatenation can be performed by embedding squares labeled by 2-fold loops into $I^2$. The relative distance and size of the squares do not matter. Note that since we can move the $2$-cubes around each other, unlike the $\EE_1$ situation in (c), one cannot assign an ordering to the multiplication. This is a reflection of the fact $\Ga_1 \circ \Ga_2 = \Ga_2 \circ \Ga_1$ for $\EE_2$-algebras.
        }
        \label{fig:E_1}
\end{figure}

Let us first discuss this structure when concatenating two 1-fold loops. As shown in Fig.~\ref{fig:E_1}a, two loops are concatenated by joining them at the basepoint $x\in X$. It is crucial that these loops end and begin at the same point $x$. This takes two 1-fold loops ${\gamma_1, \gamma_2 \in \Omega X}$ to form another 1-fold loop ${\gamma_3 = \gamma_1 \circ \gamma_2}$. Since $\ga_3\in\Om X$, loop concatenation defines a multiplication structure on the loop space ${\circ\colon  \Omega X \times \Omega X \to \Omega X}$. The unit of the multiplication is the constant path at $x$.

The algebraic structure describing 1-fold loop concatenation can be viewed as the fusion of line segments on a line. Indeed, since we consider loops up to homotopy (as in the map $\gamma\colon  I \to X$, see Fig.~\ref{fig:loops}c), ${\gamma_1 \circ \gamma_2\colon I\to X}$ does not need to be immediately concatenated in $I$. As shown in Fig.~\ref{fig:E_1}b, this allows us to deform $I$ in ${\gamma_1 \circ \gamma_2}$ such that $\ga_1$ and $\ga_2$ are disjoint line segments of $I$. The product structure is then defined as embedding line segments in $I$, each corresponding to the domain of a 1-fold loop, in the order the 1-fold loops are concatenated. For example, the loop concatenation ${\ga_1\circ\ga_2\circ\ga_3}$ is shown in Fig.~\ref{fig:E_1}c.

A benefit of this view of $\EE_1$-algebras is that it demonstrates the properties of 1-fold loop concatenation in a fairly simple way. For instance, in the example shown in Fig.~\ref{fig:E_1}c, ${\ga_1\circ\ga_2\circ\ga_3}$ can be performed by first fusing the line segments corresponding to $\ga_1$ and $\ga_2$ and then the line segment for $\ga_3$, or equivalently by first fusing the line segments corresponding to $\ga_2$ and $\ga_3$ and then the line segment for $\ga_1$. Therefore, up to homotopy,  1-fold loop concatenation satisfies a homotopic version of associativity. The two loops $(\ga_1\circ\ga_2)\circ\ga_3$ and $\ga_1\circ(\ga_2\circ\ga_3)$ are homotopy equivalent. That is, there is a homotopy between these two loops. Furthermore, in this example, the line segment $\ga_1$ can never be ``pushed through'' the line segment $\ga_2$ to change the order in which the line segments appear. Therefore, the order in which 1-fold loops concatenate matter (${\ga_1\circ\ga_2\neq \ga_2\circ \ga_1}$) and 1-fold loop concatenation is, in general, not commutative. It is useful to quantify how non-commutative a multiplication structure is. If we fix $\ga_2$ at the origin, then there are two ways to fuse $\ga_1$ into it, either on the left or the right. That is, there are $S^0$ many ways to multiply two points of an $\EE_1$-algebra, corresponding to $\ga_1 \circ \ga_2$ and $\ga_2 \circ \ga_1$.

Having explicitly discussed the structure underlying 1-fold loop concatenation, before generalizing to $n$-fold loops, let us explicitly discuss the 2-fold loop case. Similar to the 1-fold loop case, the multiplication structure of a 2-fold loop space ${\circ\colon \Om^2 X \times \Om^2 X \to \Om^2 X}$ is realized by embedding two squares in $I^2$. Given any embedding of two squares in $I^2$, we can 
concatenate two 2-fold loops ${\Ga_1,\Ga_2\in\Om^2 X}$ to produce a new 2-fold loop $\Ga_1\circ\Ga_2\in\Om^2 X$ by fusing $\Ga_1$ and $\Ga_2$ at the basepoint $x$ (see Fig.~\ref{fig:E_1}d). The algebraic structure describing 2-fold loop concatenation is that of an $\EE_2$-algebra, which can be viewed as the fusion of squares in $I^2$. More generally, given an embedding of $n$ squares in $I^2$, we can concatenate $n$ $2$-fold loops. See Fig.~\ref{fig:E_1}f for a case of ${n = 3}$.

Using this point of view of $\EE_2$-algebras, let us discuss the properties of 2-fold loop concatenation. Just as the $\EE_1$ case, ${(\Ga_1\circ\Ga_2)\circ\Ga_3 = \Ga_1\circ(\Ga_2\circ\Ga_3)}$, 2-fold loop concatenation is also (homotopically) associative. However, as opposed to 1-fold loop concatenation, the 2-fold loops ${\Ga_1\circ \Ga_2}$ and ${\Ga_2\circ\Ga_1}$ are actually (homotopy) equivalent to each other. This is because when viewed as squares in $I^2$, $\Ga_1$ and $\Ga_2$ can be moved and their places swapped. However, there are two unique ways to do this which are not necessarily equivalent: starting with ${\Ga_1\circ \Ga_2}$, $\Ga_1$ either goes over or under $\Ga_2$ to swap places. The inequivalence of these is measured by taking $\Ga_2$ and winding it around $\Ga_1$, a process which we will call braiding\footnote{Here, we have only considered $\EE_n$-algebras in topological spaces. $\EE_n$-algebras can be generalized to other structures such as categories. For example, the fusion and braiding of anyons in ${(2+1)}$D topological orders is described by an $\EE_2$-algebra in categories. Indeed, by shrinking the squares discussed here to points, they can be interpreted as anyons. While the fusion is commutative (${a\times b = b\times a}$), braiding anyon $a$ around anyon $b$ can cause the wavefunction to accumulate a nontrivial Aharonov-Bohm phase or transform nontrivially if these are non-abelian anyons.}. 

2-fold loop concatenation is only homotopically commutative. We can quantify how commutative it is by considering how many homotopy equivalent ways two 2-fold loops can be concatenated. When we fuse two $2$-cubes $\Ga_1$ and $\Ga_2$ in $I^2$, if we fix $\Ga_2$ at the origin, the different ways $\Ga_1$ fuses with $\Ga_2$ are parameterized by the angle ${\th\in [0,2\pi)}$ it approaches $\Ga_2$. Since this angle parametrizes $S^1$, we say that there are $S^1$ many ways to fuse $\Ga_1$ and $\Ga_2$. Braiding corresponds to the winding $\Ga_1$ once around $S^1$ and measures the non-commutativity of $\EE_2$-algebras.

Two particular ways of having $\Ga_1$ approach $\Ga_2$ is from ${\th = 0}$ or ${\th = \pi}$, which are the equatorial points on $S^1$. These precisely correspond to the $S^0$ many ways to multiply two points of an $\EE_1$-algebra. Since ${S^0\subset S^1}$, the fusion from ${\EE_1}$ is embedded in the fusion from ${\mb{E_2}}$. However, since $\Ga_1$ and $\Ga_2$ can exchange places in $I^2$, $\EE_2$ trivializes the $\EE_1$ non-commutativity. Thus, $\EE_2$ is more commutative than $\EE_1$.

\begin{figure}[t!]
\centering
    \includegraphics[width=.91\textwidth]{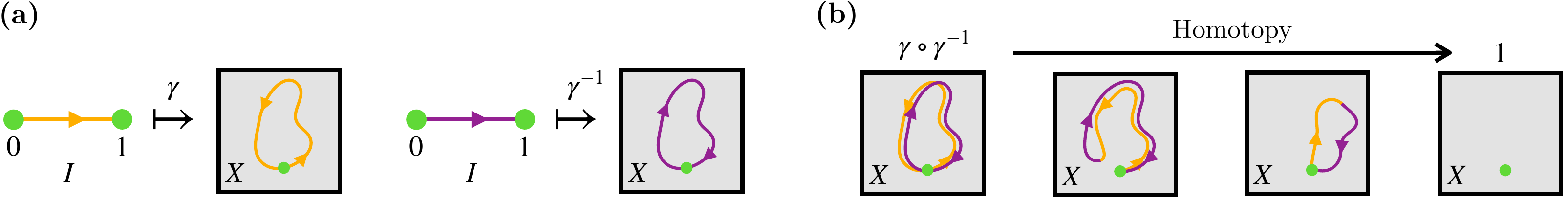}
    \caption{(a) Given a 1-fold loop $\ga$, its inverse $\ga^{-1}$ is $\ga$ with its orientation reversed.
    (b) The 1-fold loop concatenation ${\ga\circ\ga^{-1}}$ is homotopy equivalent to the constant loop $1$. Here, in the left most $X$, $\ga^{-1}$ is drawn slightly offset of $\ga$ for clarity, but in actuality $\ga^{-1}$ would be drawn on top of $\ga$.
    }
    \label{fig:loopInv}
\end{figure}

In general, $\Omega^n X$ has an $\EE_n$-algebra structure whose product corresponds to the concatenation of $n$-fold loops. Indeed, this can be seen by a straightforward generalization of the previous reasoning. Given an embedding of two $n$-cubes in $I^n$ and ${\ga_1,\ga_2\in\Om^n X}$, we can construct ${\ga_1\circ \ga_2\colon  I^n\to X}$ as follows: we map the interior of the two embedded cubes to $X$ using $\ga_1$ and $\ga_2$, and the rest of $I^n$ maps to the basepoint $x$ in $X$. More generally, given an embedding of $m$ $n$-cubes in $I^n$, we can concatenate $m$ $n$-fold loops.  

The $\EE_n$-algebra encodes the fusion of $n$-cubes in $I^n$. Furthermore, the $\EE_n$-algebra also encodes the unital structure to the multiplication: $1 \circ \gamma = \gamma = \gamma \circ 1$. The unit ${1\colon I^n\to X}$ is the constant $n$-fold loop and maps all of $I^n$ to the basepoint $x$. In the color scheme used in the figures, the constant $n$-fold loop corresponds to all of $I^n$ being colored green. 

We can also consider the ${n=0}$ case, where the fusion structure for $\EE_0$-algebras is the degenerate situation of fusions of points in $\mb{R}^0 = *$. There is no multiplication here since we can't embed two points into one; the only data is the basepoint, where we don't label any points. Therefore, an $\EE_0$-algebra is simply a space with a basepoint, such as our $(X,x)$.

$\Omega^n X$ has one more structure not encoded in $\EE_n$ fusion structure. Every point $\gamma \in \Omega^n X$ has an inverse $\gamma^{-1}$. That is, another $n$-fold loop $\gamma^{-1}$ such that ${\gamma \circ \gamma^{-1}}$ and ${\gamma^{-1} \circ \gamma}$ are homotopic to the constant $n$-fold loop $1$. For $\Omega X$, the inverse of a loop $\gamma$ is the loop of opposite orientation: $\gamma^{-1}(t) = \gamma(1-t)$, as depicted in Fig.~\ref{fig:loopInv}.  An $\EE_n$-algebra where every element has an inverse is called a grouplike $\EE_n$-algebra. In particular, $\Omega^n X$ is a grouplike $\EE_n$-algebra. For example, a group $G$ is a grouplike $\EE_1$-algebra. Furthermore, if $G$ is abelian, then it is a grouplike $\EE_\infty$-algebra.

\begin{figure}[t!]
\centering
    \includegraphics[width=.91\textwidth]{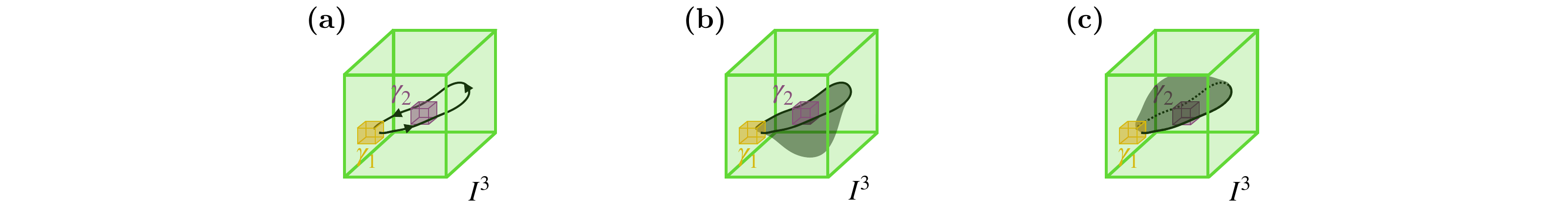}
    \caption{(a) In the concatenated 3-fold loop ${\ga_1\circ\ga_2\colon I^3\to X}$, the 3-cubes in $I^3$ which map to $\ga_1$ and $\ga_2$ can be braided. This braiding was nontrivial for $\EE_2$ but is trivialized in $\EE_3$ since the path of braiding $\ga_1$ around $\ga_2$ can be shrunk to a point by bringing it either (b) below or (c) over $\ga_2$.
    }
    \label{fig:E3Triv}
\end{figure}

Another generality is that the multiplication structure of $\EE_{n+1}$ is always more commutative than the multiplication structure of $\EE_n$. In fact, an $\EE_m$ structure is fully contained in ${\EE_n}$ for all ${0\leq m<n}$. Indeed, this was seen in our explicit treatment of $\EE_1$ and $\EE_2$. Generally, when concatenating two $n$-fold loops, there are $S^{n-1}$ many ways to fuse the two corresponding  $n$-cubes in $I^n$. Its non-commutativity is measured by a higher version of braiding over the $S^{n-1}$ many choices. Fusing the $n$-cubes through the ``equator'' of $S^{n-1}$, which is an $(n-2)$-sphere, reproduces the fusion structure of $\EE_{n-1}$-algebra. Therefore, in $\EE_n$, there are two ways to trivialize the $\EE_{n-1}$ higher braiding: either going through the northern or southern hemisphere of $S^{n-1}$ (see Fig.~\ref{fig:E3Triv} where ${n=3}$). Taking the limit $n \to \infty$ gives the $\EE_{\infty}$-algebra, which is fully commutative. 

\subsection{Maps of \texorpdfstring{$\EE_n$}{En} algebras}\label{mapsOfEnAlgs}

When we consider maps between objects endowed with algebraic structures, it is common that we would like the map to preserve this structure. A simple example is the notion of group homomorphisms between groups. Given two groups $G$ and $H$, whose group operations are denoted as $*$ and $\bm{\cdot}$, respectively, we could consider maps ${F\colon  G \to H}$ that preserve the group structure. That is, given ${g_1, g_2 \in G}$, we want $F$ to satisfy:
\begin{equation}
F(g_1 * g_2) = F(g_1) \bm{\cdot} F(g_2),
\end{equation}
where ${F(g_1),F(g_2)\in H}$. 

Having introduced $\Om^n X$ and its grouplike $\EE_n$-algebra structure in the previous two subsections, it is natural to consider maps preserving the $\EE_n$ structure. In fact, since a group is an example of a grouplike $\EE_1$-algebra, the above discussion on group homomorphisms is an example of maps preserving an $\EE_1$ structure. Another familiar example is based maps between based spaces. Indeed, since an $\EE_0$-algebra is just a space with a basepoint, a map preserving the ${\EE_0}$ structure of $(X,x)$ and $(Y,y)$ is ${F\colon  X \to Y}$ such that $F(x) = y$. Generally, given two $\EE_n$-algebras $Y$ and $Y'$, we can consider the map ${F\colon  Y\to Y'}$ that preserves the $\EE_n$ structure. For instance, $Y$ and $Y'$ could both be $n$-fold loop spaces, and the fact that $F$ preserves the $\EE_n$ structure means that $F$ preserves the $n$-fold loop concatenation structure of $Y$. To emphasize that $F$ preserves this $\EE_n$ structure, we say that $F$ is a map of $\EE_n$-algebras.\footnote{Note that additional homotopy data is needed to specify a map of $\EE_n$-algebras. Like spin structures on manifolds, there can be multiple inequivalent choices.}

Given a map between spaces $A$ and $B$, one often wants to consider the set of maps up to homotopy, denoted as $[A, B]$. This can also be done with maps that preserve an algebraic structure. Indeed, we can also classify homotopy classes of maps that preserve some algebraic structure, like the $\EE_n$-algebra structures. The homotopy class of $\EE_n$-algebra maps between $\EE_n$-algebras $Y$ and $Y'$ is denoted as 
\begin{equation}
     [Y, Y']_{\EE_n}.
\end{equation}

Given given two $\EE_0$-algebras ${(X,x)}$ and ${(X',x')}$ and an $\EE_0$-algebra map ${F\colon  X \to X'}$ (i.e., ${F(x) = x'}$), $F$ takes $n$-fold loops in $(X,x)$ to $n$-fold loops in $(X',x')$. That is, from $F$ we can define a map ${\Omega^n F\colon  \Omega^n X \to \Omega^n X'}$. Furthermore, this respects the $\EE_n$-algebra structure: give two $n$-fold loops ${\gamma_1, \gamma_2 \in \Omega^n X}$, then $F(\gamma_1 \circ_X \gamma_2) = F(\gamma_1) \circ_{X'} F(\gamma_2)$, where $\circ_X$ and $\circ_X'$ are the concatenation operations of $\Om^n X$ and $\Om^n X'$, respectively. Therefore $\Om^n F$ is an $\EE_n$-algebra maps. On the level of homotopy classes, the construction ${F \mapsto \Omega^n F}$ induces a map
\begin{equation}\label{E0toEnmaphomotopy}
    [X, X']_{\EE_0} \to [\Omega^n X, \Omega^n X']_{\EE_n}.
\end{equation}

\subsection{Deloop}\label{deloopAppendix}

After the previous two subsections on $\EE_n$-algebras, we can now return to loop spaces with a new point of view. Recall that a space $X$ with a basepoint is an $\EE_0$-algebra while the loop space $\Om X$ is a grouplike $\EE_1$-algebra. Therefore, the loop space construction takes an $\EE_0$-algebra and returns a grouplike $\EE_1$-algebra:
\begin{equation*}
    \begin{tikzcd}
        \left\{\text{$\EE_0$-algebras}\right\} \arrow{r}{\Om}
        &
        \left\{\text{Grouplike $\EE_1$-algebras}\right\} .
    \end{tikzcd}
    \end{equation*}
There is an inverse construction, called deloop, that takes a grouplike $\EE_1$-algebra $Y$ to an $\EE_0$-algebra $BY$, such that we can recover $Y$ from $BY$:
\begin{equation}\label{eq:LoopDeloop}
    Y \simeq \Omega BY,
\end{equation}
where $\simeq$ means homotopy equivalence. 

Eq.~\eqref{eq:LoopDeloop} implies that any grouplike $\EE_1$-algebra is homotopic equivalent to a loop space and that any point of ${y \in Y}$ corresponds to a based loop ${\gamma_y}$ in ${BY}$. Furthermore, since $Y$ is a grouplike $\EE_1$-algebra, it has an associative product structure ${*\colon  Y \times Y \to Y}$, and the equivalence of the grouplike $\EE_1$-algebras in Eq.~\eqref{eq:LoopDeloop} means that the multiplication in $Y$ corresponds to loop concatenation in $BY$:
\begin{equation}
    \gamma_{y * y'} = \gamma_y \circ \gamma_{y'}.
\end{equation}
Using Eq.~\eqref{eq:nHomotopy1loop} with ${X = BY}$ yields the expression ${\pi_i(\Omega BY) = \pi_{i+1} (BY)}$. Thus, in light of Eq.~\eqref{eq:LoopDeloop}, the homotopy groups of $BY$ are shifted up by one from the homotopy groups of $Y$:
\begin{equation}\label{homotopyBY}
    \pi_{i}(BY) = \pi_{i-1}(Y),\quad\quad \pi_0(BY) = 0.
\end{equation}
$BY$ can generally be constructed from $Y$ using simplicial methods~\cite{May06}, which we will not review here. 

It is natural to wonder if a general $\EE_0$ algebra $(X,x)$ satisfies $X \simeq B\Omega X$. On one hand, Eq.~\ref{homotopyBY} implies that 
\begin{equation}
    \pi_i(B\Omega X)) = \pi_{i-1}(\Omega X) = \pi_{i}(X) \quad\quad (i > 0),
\end{equation}
so the higher homotopy groups are the same. However, while $BY$ is connected, $X$ might not be, so their zeroth homotopy groups can differ. Indeed, for a general based space $(X,x)$, the loop space $\Om X$ cannot determine the disconnected components of $X$ because any loop based at ${x \in X}$ can only explore the path component of $x$ in $X$. For this reason, we see that $X$ is homotopically equivalent to $B\Omega X$ if and only if $X$ is connected, i.e., $\pi_0(X) = 0$.

\begin{figure}[t!]
    \centering
    \includegraphics[width=.91\textwidth]{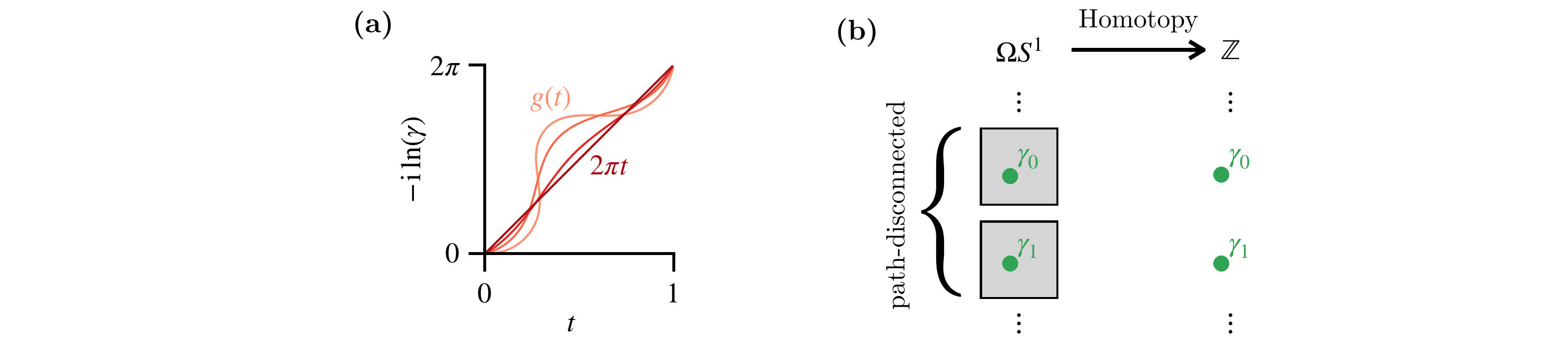}
        \caption{(a) Given a 1-fold loop ${\gamma(t) = \ee^{\ii g(t)}}$ that has winding number one, there is a homotopy to the standard constant velocity loop ${\gamma_1(t) = e^{2 \pi i t}}$. This can be done for a general 1-fold loop with any winding number. (b) The space $\Omega S^1$ has $\mb{Z}$ components labeled by the winding number. While this space is huge, each component is contractible using the contraction above. Therefore, $\Omega S^1$ is homotopy equivalent to $\mb{Z}$.}
        \label{fig:last}
\end{figure}

Let us consider the example of ${Y = \mb{Z}}$ with addition as the grouplike $\EE_1$-algebra structure. From the above discussion, there exists a path-connected space ${BY\equiv B\Z}$ such that ${\Omega B\mb{Z} \simeq \mb{Z}}$. Since ${\pi_0(\Z) = \Z}$ while ${\pi_i(\Z) = 0}$ for all ${i\neq 0}$, from Eq.~\eqref{homotopyBY} the homotopy groups of $B\Z$ must satisfy
\begin{equation}
    \pi_i(B\mb{Z}) = 
    \begin{cases}
    \Z\quad &\txt{if }i = 1,\\
    0~\quad&\txt{if }i\neq 1.
    \end{cases}
\end{equation}
It's clear that $S^1$ fits these criteria, and thus ${B\Z = S^1}$. Now let's check that ${\Om S^1 \simeq \mb{Z}}$. For each $n \in \mb{Z}$, we have a constant velocity loop 
\begin{equation}
    \gamma_n(t) = e^{2\pi \ii nt},
\end{equation}
where ${t\in[0,1]}$. The loop $\gamma_n(t)$ has winding number $n \in \mb{Z}$. It is clear that for $n \neq m$, $\gamma_n$ and $\gamma_m$ are not homotopic since the winding number is a discrete invariant that doesn't change under homotopy. Furthermore, any loop $\gamma \in \Om S^1$ with winding number $n$ is homotopy equivalent to $\gamma_n$. This is because we can write $\gamma(t) = e^{2 \pi \ii g(t)}$, with $g(t)$ being a function from $[0,1] \to \mb{R}$ with boundary condition $g(0) = 0$, $g(1) = n$. We can continuously morph $g$ into the linear function $g_n(t) = nt$ (see Fig ~\ref{fig:last} (a) for the case $n = 1$). This defines a homotopy between $\gamma$ and $\gamma_n$. This means that $\Om S^1$ has $\mb{Z}$ components, which label the winding number. Furthermore, each component is contractible, meaning that they can be continuously deformed to a single point (see Fig ~\ref{fig:last} (b)). 

Furthermore, this homotopy equivalence preserves the grouplike $\EE_1$-algebra structure: given two loops that wind around $n_1$ and $n_2$ times, respectively, their concatenation winds around $n_1 + n_2$ times. Therefore, this shows that $\Omega S^1 \simeq \mb{Z}$ as a grouplike $\EE_1$-algebra.

The above can be straightforwardly generalized to $\Om^n$. Constructing the $n$-fold loop space $\Om^n X$ can be viewed as taking an $\EE_0$-algebra $X$ to the grouplike $\EE_n$-algebra $\Om^n X$:
\begin{equation*}
    \begin{tikzcd}
        \left\{\text{$\EE_0$-algebras}\right\} \arrow{r}{\Om^n}
        &
        \left\{\text{Grouplike $\EE_n$-algebras}\right\} .
    \end{tikzcd}
\end{equation*}
The inverse construction takes a grouplike $\EE_n$-algebra $Y$ to an $\EE_0$-algebra ${B^n Y}$ such that 
\begin{equation}\label{BnOmnY}
    \Omega^n B^n Y  \simeq Y
\end{equation}
as grouplike $\EE_n$-algebras. From the fact that ${\pi_i(\Omega^n X) = \pi_{i + n}(X)}$ (see Eq.~\eqref{nLoopkHomotopyraising}), $B^n Y$ satisfies
\begin{equation}
        \pi_i(B^n Y) = \begin{cases} \pi_{i-n}(Y)\quad\quad &i \geq n,\\
         0 \quad \quad &i < n.
        \end{cases}
\end{equation}
Therefore, $B^n$ shifts the homotopy groups down by $n$.  Recall that a space $X$ is $n$-connected if its first $n$ homotopy groups are $0$. We see that $B^n Y$ is $n$-connected where $Y$ is a grouplike $\EE_n$-algebra. Furthermore, generalizing the $n = 1$ case, this implies that if $X$ is a $n$-connected space (such as $S^k$ for $k \geq n$), $X \simeq B^n \Omega^n X$.

Lastly, recall that the loop space construction was iterative. This means that the ${p+q}$ loops space satisfies the relation ${\Omega^p(\Omega^q(X)) = \Omega^{p+q}(X)}$. This is also true for delooping constructions. That is, 
\begin{equation}
    B^p(B^q Y) = B^{p+q}Y.
\end{equation}
Proving this requires the fact that for ${q < n}$, the $q$-delooping of an $\EE_n$-algebra $Y$, denoted as ${B^q Y}$, is an $\EE_{n-q}$-algebra. The interested reader can refer to Ref.~\citenum{May06} for a more detailed discussion on iterated-delooping.

\subsection{Duality between loops and delooping}\label{loopDeloopDuality}

The looping construction takes $\EE_0$-algebras to grouplike $\EE_n$-algebras while the delooping construction takes grouplike $\EE_n$-algebras to $n$-connected $\EE_0$-algebras. Since $n$-connected $\EE_0$-algebras are a subset of all $\EE_0$-algebras, we therefore have the relation
\begin{equation*}
\begin{tikzcd}
    \hspace{-4pt}\left\{\text{$n$-connected $\EE_0$-algebra}\right\} \hspace{-2pt}\arrow[shift left=.5ex]{r}{\Om^n}
    &
    \hspace{-2pt}\left\{\text{Grouplike $\EE_n$-algebra}\right\} 
    \arrow[shift left=.5ex]{l}{B^n}\hspace{-1pt}.
\end{tikzcd}
\end{equation*}
In fact, as discussed above, the left-hand side contains the same data as the right-hand side. Therefore, $n$-connected $\EE_0$-algebras and grouplike $\EE_n$-algebras are dual to one another, and $B^n$ and $\Omega^n$ translate between these two dual descriptions.\footnote{We note that $\EE_n$-algebras and $\EE_n$-algebra maps can be packaged together into the structure of an $\infty$-category. In fact, there is a mathematical theorem first proven by May (see  \cite[Thm 1.4]{May06}) that states $\Omega$ and $B$ give an equivalence of these categories. We also refer the reader to Ref.~\cite[Thm 5.2.6.10]{HA}.} 

This duality between grouplike $\EE_n$-algebras and $n$-connected $\EE_0$-algebras extends to a duality of maps. For instance, returning to Eq.~\eqref{E0toEnmaphomotopy}, this implies that when $X$ and $X'$ are $n$-connected, the construction ${F\mapsto\Om^n F}$ induces an equivalence
\begin{equation}\label{E0toEnmaphomotopy2}
    [X, X']_{\EE_0} \simeq [\Omega^n X, \Omega^n X']_{\EE_n}.
\end{equation}
Furthermore, letting $X$ and $X'$ in Eq.~\eqref{E0toEnmaphomotopy2} be the $n$-connected $\EE_0$-algebras $B^n Y$ and $B^n Y'$, where $Y$ and $Y'$ are grouplike $\EE_n$-algebras, we find 
\begin{equation}\label{eq:another-one}
    [B^n Y, B^n Y']_{\EE_0} \simeq [\Omega^n B^n Y, \Omega^n B^n Y']_{\EE_n}\simeq [Y, Y']_{\EE_n} ,
\end{equation}
where the last step follows from Eq.~\eqref{BnOmnY}.

Let us consider the ${n=1}$ case explicitly in the example ${Y = Y' = \mb{Z}}$. Then an $\EE_1$-algebra map ${F\colon  \mb{Z} \to \mb{Z}}$ is determined by where $1\in Y$ is sent to in $Y'$. In particular, if ${F(1) = m}$, where ${m\in Y'}$, then for all ${n\in Y}$, ${F(n) = nm}$. Therefore there are $\mb{Z}$ worth of maps that preserves the $\EE_1$ structure, and so 
\begin{equation}\label{exZE1}
    [\mb{Z}, \mb{Z}]_{\EE_1} = \mb{Z}.
\end{equation}
As we saw in the previous subsection, $\Z$ with an addition structure is a grouplike $\EE_1$-algebra that satisfies ${B\Z \simeq S^1}$. Using this, Eq.~\eqref{eq:another-one} claims that 
\begin{equation}\label{exZE1toE0}
    [\Z,\Z]_{\EE_1} \simeq [S^1,S^1]_{\EE_0}.
\end{equation}
This is easily verified. Indeed, the right-hand side is simply
\begin{equation}
    [S^1, S^1]_{\EE_0} \equiv \pi_1(S^1) = \mb{Z}.
\end{equation}
And thus, in light of Eq.~\eqref{exZE1}, Eq.~\eqref{exZE1toE0} is indeed satisfied.

\section{Calculation of \texorpdfstring{$[\Omega^n S^k, U(1)]_{\EE_n}$}{[Omega Sk, U(1)]EEn}}\label{U(1)HomotopyTheoryForFunAppendix}

In this appendix section, we compute the homotopy class
\begin{equation}\label{eq:C-1}
    [\Omega^n S^k, U(1)]_{\EE_n}.
\end{equation}
Since we have already computed the $k\geq n$ case in the main text (see Eq.~\eqref{loopSphereOP3}), we need only need to consider ${k< n}$. Throughout this calculation, we will assume familiarity with algebraic topology, particularly spectra. We refer the reader to Ref.~\citenum{beaudry2018guide} for an introduction for physicists.

The ${n = 1}$ and $2$ cases for ${k<n}$ are straightforward. Indeed, since $\Omega^1 S^0 = \Omega^2 S^0 = \Omega^2 S^1 = *$, Eq~\eqref{eq:C-1} is trivial for these cases. So, using Eq.~\eqref{loopSphereOP3}, we get the classification for ${n\leq 2}$:
\begin{equation}\label{appendCnleq2}
    [\Omega^n S^k, U(1)]_{\EE_n} = \begin{cases}
        \Z\quad\quad & k = n+1,\\
        0 \quad\quad &\txt{else}.\\
    \end{cases}\quad\quad\quad\quad (n\leq 2)
\end{equation}

To calculate the homotopy classes when ${n \geq 3}$, we translate Eq.~\eqref{eq:C-1} into a question in stable homotopy theory. Since $U(1)$ is a $1$-truncated space (i.e., ${\pi_n(U(1)) = 0}$ for ${n > 2}$), we can rewrite Eq.~\eqref{eq:C-1} as
\begin{equation}\label{eq:C-2}
    [\tau_{\leq 1} \Omega^n S^k, U(1)]_{\EE_n} = [\tau_{\leq 1} \Omega^n S^k, U(1)]_{\EE_\infty},
\end{equation}
where ${\tau_{\leq 1} \Omega^n S^k}$ is the 1st Postnikov stage of $\Omega^n S^k$. The equality follows from the fact that $\EE_n$-algebra structures are equivalent to $\EE_\infty$-algebra structures for $1$-truncated spaces. 

Since grouplike $\EE_\infty$-algebras are equivalent to connective spectra, we can find spectra corresponding to $U(1)$ and $\tau_{\leq 1} \Omega^n S^k$. The spectrum associated to $U(1)$ is $\Sigma H\mathbb{Z}$, where $H\mathbb{Z}$ is the Eilenberg-MacLane spectrum associated to $\mathbb{Z}$. On the other hand, the spectrum $\mathcal{X}$ associated to 
$\tau_{\leq 1} \Omega^n S^k$ is a $1$-truncated connective spectrum (i.e., it only has homotopy groups in degree $0$ and $1$), satisfying 
\begin{equation}\label{eq:X-hom-groups}
    \pi_0(\mathcal{X}) = \pi_n(S^k), \quad\quad\quad\quad \pi_1(\mathcal{X}) = \pi_{n+1}(S^k).
\end{equation}
Furthermore, its $k$-invariant ${k \colon H(\pi_{n} (S^k)) \to \Sigma^2 H(\pi_{n+1}(S^k))}$  is given by the $k$-invariant of $S^k$ connecting ${\pi_n(S^k)}$ and ${\pi_{n+1}(S^k)}$. Therefore, Eq.~\eqref{eq:X-hom-groups} gives us a fiber sequence
\begin{equation}\label{eq:C-4}
\Sigma H(\pi_{n+1}(S^k)) \to \mathcal{X} \to H(\pi_{n}(S^k)).
\end{equation}

We can now translate Eq.~\eqref{eq:C-2} to a statement in spectra. Indeed, using the above
\begin{equation}\label{eq:C-5}
    [\tau_{\leq 1} \Omega^n S^k, U(1)]_{\EE_\infty} = [\mathcal{X}, \Sigma H\mathbb{Z}] = H\mathbb{Z}^1(\mathcal{X}),
\end{equation}
and the fiber sequence Eq.~\eqref{eq:C-4} gives a long exact sequence:
\begin{equation}\label{eq:C-6}
    0 \to H\mathbb{Z}^1(\pi_{n}(S^k))
    \to H\mathbb{Z}^1(\mathcal{X}) \to H\mathbb{Z}^0(H\pi_{n+1}(S^k)) \to H\mb{Z}^2(H\pi_n(S^k)) \to \cdots.
\end{equation}
We can deduce ${H\mathbb{Z}^1(\mathcal{X})}$, and therefore ${[\Omega^n S^k, U(1)]_{\EE_n}}$, using this long exact sequence.

To start, we have 
\begin{equation}
    H\mathbb{Z}^1(\pi_{n}(S^k)) \simeq \txt{Hom}(\pi_{n}(S^k)^{\txt{tor}}, U(1))
    \end{equation}
and 
\begin{equation}
    H\mathbb{Z}^0(H\pi_{n+1}(S^k)) = \txt{Hom}(\pi_{n+1}(S^k)^{\txt{free}}, \mathbb{Z}),
\end{equation}
where ${\pi_{n}(S^k)^{\txt{tor}}}$ is the torsion subgroup of ${\pi_n(S^k)}$, and ${\pi_{n+1}(S^k)^{\txt{free}} \hspace{-2pt}= \hspace{-2pt} \pi_{n+1}( \hspace{-1pt}S^k )/\pi_{n+1}( \hspace{-1pt}S^k)^{\txt{tor}}}$ is the free quotient. Note that ${\txt{Hom}(\pi_{n}(S^k)^{\txt{tor}}, U(1))}$ is the Pontryagin dual of ${\pi_{n}(S^k)^{\txt{tor}}}$ and is abstractly isomorphic to $\pi_{n}(S^k)^{\txt{tor}}$. On the other hand, ${\txt{Hom}(\pi_{n+1}(S^k)^{\txt{free}}, \mathbb{Z})}$ is the lattice dual to ${\pi_{n+1}(S^k)^{\txt{free}}}$ and is also abstractly isomorphic to ${\pi_{n+1}(S^k)^{\txt{free}}}$. 

Furthermore, since ${H\mb{Z}^2(H\pi_n(S^k))}$ is a torsion group, by Eq.~\eqref{eq:C-6} we get a short exact sequence:
\begin{equation}\label{eq:C-7}
   0 \to  \txt{Hom}(\pi_{n}(S^k)^{\txt{tor}}, U(1)) \to   H\mathbb{Z}^1(\pi_{n}(S^k)) \to A \to 0
\end{equation}
where ${A = \ker(H\mathbb{Z}^0(H\pi_{n+1}(S^k)) \to H\mb{Z}^2(H\pi_n(S^k))}$ is a free abelian group of the same rank as $\pi_{n+1}(S^k)^{\txt{free}}$. Since $A$ is free,  Eq.~\eqref{eq:C-7} splits and 
\begin{equation}
    H\mathbb{Z}^1(\pi_{n}(S^k)) \simeq \txt{Hom}(\pi_{n}(S^k)^{\txt{tor}}, U(1)) \oplus A.
\end{equation}

Note that $\pi_{n+1}(S^k)^{\txt{free}}$ is non-trivial if and only if ${n = k-1}$ or ${n = 2k-2}$ and ${k \in 2\mathbb{Z}}$, where ${\pi_{n+1}(S^k)^{\txt{free}} = \mathbb{Z}}$. This implies that ${A = \mathbb{Z}}$ when ${n = k-1}$ or ${n = 2k-2}$ and ${k \in 2\mathbb{Z}}$, and $0$ otherwise. Putting it all together, in the case that $n \geq 3$, we get 
\begin{align}\label{appendCn3}
     [\Omega^n S^k, U(1)]_{\EE_n} &=  H\mathbb{Z}^1(\mathcal{X}) \quad\quad\quad\quad (n\geq 3)\nonumber\\
     &= 
     \begin{cases}
             0 \quad \quad & n < k-1,\\
        \mathbb{Z} \quad\quad &n = k-1, \\ 
         \hom(\pi_{n}(S^k), U(1)) \oplus \mathbb{Z} \quad\quad &n = 2k -2, k \in 2 \mathbb{Z}, k > 2 \\ 
          \hom(\pi_{n}(S^k)^{\txt{tor}}, U(1)) \quad\quad &\txt{else}.
     \end{cases}
\end{align}

Combining Eq.~\eqref{appendCnleq2} and~\eqref{appendCn3}, we get that 
\begin{equation}
     [\Omega^n S^k, U(1)]_{\EE_n} =  
     \begin{cases}
             0 \quad \quad & n < k-1,\\
        \mathbb{Z} \quad\quad &n = k-1, \\ 
        0 \quad \quad & n = 2, k = 2\\
         \hom(\pi_{n}(S^k), U(1)) \oplus \mathbb{Z} \quad\quad &n = 2k -2, k \in 2 \mathbb{Z}, k > 2 \\ 
          \hom(\pi_{n}(S^k)^{\txt{tor}}, U(1)) \quad\quad &\txt{else}.
     \end{cases}
\end{equation}
As mentioned above, since $\pi_{n}(S^k)^{\txt{tor}}$ is a finite abelian group, its Pontryagin dual is abstractly isomorphic to itself, so we can rewrite this as 
\begin{equation}
     [\Omega^n S^k, U(1)]_{\EE_n} =  
     \begin{cases}
             0 \quad \quad & n < k-1,\\
        \mathbb{Z} \quad\quad &n = k-1, \\ 
        0 \quad \quad & n = 2, k = 2\\
         \pi_{n}(S^k)\oplus \mathbb{Z} \quad\quad &n = 2k -2, k \in 2 \mathbb{Z}, k > 2 \\ 
          \pi_{n}(S^k)^{\txt{tor}} \quad\quad &\txt{else}.
     \end{cases}
\end{equation}

\end{appendix}

\bibliography{refs.bib} 

\end{document}